# Study of Parameterized-Chain Networks

M. H. Zibaeenejad and J. G. Thistle*

**Abstract**

In areas such as computer software and hardware, manufacturing systems, and transportation, engineers encounter networks with arbitrarily large numbers of isomorphic subprocesses. Parameterized systems provide a framework for modeling such networks. The analysis of parameterized systems is a challenge as some key properties such as nonblocking and deadlock-freedom are undecidable even for the case of a parameterized system with ring topology. In this paper, we introduce *Parameterized-Chain Networks* (PCN) for modeling of networks containing several linear parameterized segments. Since deadlock analysis is undecidable, to achieve a tractable subproblem we limit the behavior of subprocesses of the network using our previously developed mathematical notion 'weak invariant simulation.' We develop a *dependency graph* for analysis of PCN and show that partial and total deadlocks of the proposed PCN are characterized by *full, consistent* subgraphs of the dependency graph. We investigate deadlock in a traffic network as an illustrative example.

## I. Introduction

A *parameterized network* is composed of arbitrary finite numbers of isomorphic subprocesses. Formally, such systems can be modeled as infinite families of finite-state systems. They are a subclass of the so-called 'parameterized systems', whose models incorporate parameters with unspecified values [1]. In the case of parameterized networks, the parameter is the number of subprocesses in the network. Practical examples of parameterized networks include wireless sensor networks, transportation networks, manufacturing systems and subprocesses in operating systems. Parameterized models are particularly useful when the number of subprocesses is unknown, time-varying, or very large.

It is natural to ask how much analysis and control can be done independently of a specific parameter values. Unfortunately, key problems such as checking the nonblocking property for parameterized networks are generally undecidable [2]. Parameterized networks have received considerable attention in the model-checking literature [3], [4]. Most recently, the authors of [5] seek to determine whether or not a given safety property holds for all instances of parameterized toroidal mesh networks under process symmetry assumptions.

*M. H. Zibaeenejad and J. G. Thistle are with the Department of Electrical and Computer Engineering, University of Waterloo, Waterloo, ON, Canada. `mhzibaee,jthistle@uwaterloo.ca`

Within control literature, the deadlock analysis of a class of parameterized networks was considered, where subsystems are identical and interact only via events that are shared with all other subsystems [6]. This requires the communication topology of network to be that of a graph-theoretic clique. In previous work [7], the present authors introduced a novel mathematical tool, *weak invariant simulation*, to support deadlock analysis of parameterized networks. Although the deadlock-freedom property is generally undecidable in ring networks, weak invariant simulation relations was used to define a class of these networks in which all the reachable deadlocked states can be calculated [2]. In this paper, we consider Parameterized-Chain Networks (PCN) consisting of multiple linear parameterized segments together with a finite number of finite-state subprocesses having arbitrary structure.

In networks consisting of several subprocesses, nontrivial deadlocks often occur in the presence of a circular wait. When a circular wait occurs, the only available action of each subprocess requires a resource that is being held by another subprocess [8], [9]. Graph-theoretic techniques are used to characterize such dependencies in finite-state systems [9], [10]. Unfortunately, these techniques are not directly applicable to the analysis of parameterized networks.

In this paper, we characterize dependencies between subprocesses of any instances of a PCN by means of a single, finite *dependency graph*. In a preliminary form, the dependency graph was introduced in [11], where it was conjectured it can be used to detect reachable partial deadlocks of a PCN. Here we prove that specific subgraphs of the dependency graph *represent* reachable generalized circular waits of instances of the PCN. We relate partial and total deadlocks of the PCN to these generalized circular waits. Specifically, we show that the existence of a generalized circular wait is a necessary condition for total deadlock and a sufficient condition for partial deadlock of all but an acyclic subgraph of a PCN. In some applications this yields a necessary and sufficient condition for total deadlock. We illustrate our proposed method by analysis of a traffic network.

Section II covers preliminaries. Section III introduces PCN and a running example of a train network. Section IV presents our deadlock analysis method. Section V expresses the main results of the paper: the deadlock analysis of PCN by computation of the set of reachable generalized circular waits using dependency graphs. Finally, Section VI summarizes the results.

## II. Preliminaries

### A. Graphs

For the purposes of this paper, a directed graph $\mathscr{D}$ is an ordered pair $(V, A)$, where $V$ is the node set and $A$ is a set of ordered pairs of nodes called arcs. Considering an arc $(u_1, u_2)$, $u_2$ is a direct successor of $u_1$, and $u_1$ is a direct predecessor of $u_2$; the arc is an incoming arc of $u_2$ and outgoing arc of $u_1$. The number of incoming arcs to a node is called its in-degree, and the number of outgoing arcs from a node is called the out-degree of that node. A directed graph $\mathscr{D}$ is strongly connected if for every pair $u, v \in V$, $\mathscr{D}$ contains sequences of arcs linking $u$ to $v$. A closed walk is a sequence of nodes starting and ending at the same node, with each two consecutive nodes in the sequence adjacent to each other in the graph. A simple circuit is a closed walk with no repetitions of nodes, other than the repetition of the starting and ending node. For more on graph theory, see [12].

### B. Discrete event systems basics

One of the conventional ways of presenting a DES employs *generators* [13]. In this paper, the terms (sub)processes and generators are used interchangeably. A nondeterministic generator is formally defined as a 4-tuple $G = (X, \Sigma, \xi, x^0)$, where $X$ is a state set, $\Sigma$ a finite alphabet representing a finite event set, $\xi : X \times \Sigma \to 2^X$ is a transition function (where $2^X$ is the power set of $X$), and $x^0$ an initial state.[1] When $\xi(x, \sigma) \neq \emptyset$ we say that the transition $\xi(x, \sigma)$ is defined or enabled. We denote by $\Sigma^+$ the set of all nonempty finite strings of events in $\Sigma$, and $\Sigma^* = \Sigma^+ \cup \{\epsilon\}$, where $\epsilon$ denotes the empty string (the identity element for string concatenation). The transition function extends to $\xi : X \times \Sigma^* \to 2^X$ in a standard manner [13]. A shared event between two generators is an event that is enabled from states of these generators. It can occur if both of the generators are in states that allow the shared event: transitions labeled by a shared event occur simultaneously in generators that share the event. Local events are not shared with any other generator. The semantics of shared and local events are formalized by means of *synchronous products*. The synchronous product $G_1 \| G_2$ of generators $G_i = (X_i, \Sigma_i, \xi_i, x_i^0)$, $i \in \{1, 2\}$ is the reachable component of $((X_1 \times X_2), \Sigma_1 \cup \Sigma_2, \xi, (x_1^0, x_2^0))$, where

$$\xi((x_1, x_2), \sigma) = \begin{cases} \xi_1(x_1, \sigma) \times \xi_2(x_2, \sigma), & \text{if } \sigma \in \Sigma_1 \cap \Sigma_2; \\ \xi_1(x_1, \sigma) \times \{x_2\}, & \text{if } \sigma \in \Sigma_1 \setminus \Sigma_2; \\ \{x_1\} \times \xi_2(x_2, \sigma), & \text{if } \sigma \in \Sigma_2 \setminus \Sigma_1; \\ \emptyset, & \text{otherwise.} \end{cases}$$

---

[1] We write 0 as a superscript because we reserve subscripts on state symbols to represent components of tuples of states.

The definition extends naturally to $M \geq 2$ generators.

The *natural projection* [13] is defined as $P_{\hat{\Sigma}} : \Sigma^* \to \hat{\Sigma}^*$, such that

$$P_{\hat{\Sigma}}(\epsilon) = \epsilon; \quad P_{\hat{\Sigma}}(\alpha) = \begin{cases} \alpha, & \text{if } \alpha \in \hat{\Sigma}; \\ \epsilon, & \text{if } \alpha \notin \hat{\Sigma}; \end{cases}$$

$$P_{\hat{\Sigma}}(s\alpha) = P_{\hat{\Sigma}}(s) P_{\hat{\Sigma}}(\alpha), \text{ for all } s \in \Sigma^*, \ \alpha \in \Sigma.$$

For a synchronous product $\|_{i=1}^{M} G_i$, with $G_i = (X_i, \Sigma_i, \xi_i, x_i^0)$, $M \in \mathbb{N}$, and a shared event $\sigma_i$ of subprocess $G_i$, $1 \leq i \leq M$, *companion states* of $\sigma_i$ in subprocess $G_j$, $1 \leq j \leq M$ are states $x_j$, for which $\xi_j(x_j, \sigma_i) \neq \emptyset$. The set of such companion states is $\chi_j(\sigma_i)$ [2].

## C. Weak invariant simulation

We first define *weak simulation* and then *weak invariant simulation*. Consider generators $G_i = (X_i, \Sigma_i, \xi_i, x_i^0)$, $i \in \{1, 2\}$, and a natural projection $P_{\hat{\Sigma}} : \Sigma^* \to \hat{\Sigma}^*$ with $\hat{\Sigma} \subseteq \Sigma$, $\Sigma = \Sigma_1 \cup \Sigma_2$.

**Definition 1.** [14] A *weak simulation* of $G_2$ by $G_1$ with respect to $\hat{\Sigma}$ is a binary relation $\mathcal{WS} \subseteq X_1 \times X_2$ between states of the two generators $G_1$ and $G_2$ such that for each $(x_1, x_2) \in \mathcal{WS}$ and every $l_2 \in \Sigma_2^*$, if $x_2' \in \xi_2(x_2, l_2) \neq \emptyset$, there exists $l_1 \in \Sigma_1^*$ and $x_1' \in \xi_1(x_1, l_1)$ such that $P_{\hat{\Sigma}}(l_1) = P_{\hat{\Sigma}}(l_2)$ and $(x_1', x_2') \in \mathcal{WS}$. [2]

---

[2]Simulations of $G_2$ by $G_1$ are often defined elsewhere in the literature as subsets of $X_2 \times X_1$.

**Definition 2.** [2] Let $\mathcal{I}$ be a weak simulation relation of $G_2$ by $G_1$ with respect to $\hat{\Sigma}$. The weak simulation relation $\mathcal{I}$ is a *weak invariant simulation* w.r.t. $\hat{\Sigma}$ if for any pair $(x_1, x_2) \in \mathcal{I}$ and for all $l_1 \in \Sigma_1^*$, $l_2 \in \Sigma_2^*$ and all $x_1' \in \xi_1(x_1, l_1)$, $x_2' \in \xi_2(x_2, l_2)$, we have

$$P_{\hat{\Sigma}}(l_1) = P_{\hat{\Sigma}}(l_2) \Rightarrow (x_1', x_2') \in \mathcal{I}.$$

For more on weak invariant simulation, see [7], [2].

### III. The network model

#### A. An illustrative example: traffic network

Before we present our framework, we bring in a running example. Consider the train traffic network of Figure 1(a) with two intersections and three routes with arbitrary lengths. In PCN modeling of this network, intersections are distinguished subprocesses (their structure is dissimilar to the rest of the network) and routes are parameterized segments. We assume that each train entering the network consists of two cars; hence a train occupies two spaces of each route (one for each car). Each intersection will accommodate exactly one train at a time; no new train is allowed in the intersection until the previous one completely leaves. Other interesting variants of this problem can be obtained by considering more complex network structures and trains with different numbers of cars.

In the traffic network of Figure 1(a), trains enter the network from intersection one and continue to the main route. When a train arrives at intersection two, it decides to leave the network or to turn onto one of the branches. Consider an instance of the network where the main, top, and bottom routes have lengths 20, 12, and 17 respectively. We model the last two spaces of the top and bottom routes by distinguished subprocesses; therefore the parameterized segments $R'$, and $R''$ respectively contain 10 and 15 subprocesses in this instance of the network (see Figure 1(b),(c)). We will present the deadlock analysis of the parameterized network, where the routes have arbitrary lengths, after the description of our results.

#### B. Linear parameterized discrete event systems

For the purposes of this paper, a Parameterized Discrete Event System(PDES) $\mathcal{P}$ is an infinite set of synchronous products of $M$ isomorphic finite-state subprocesses, where $M$

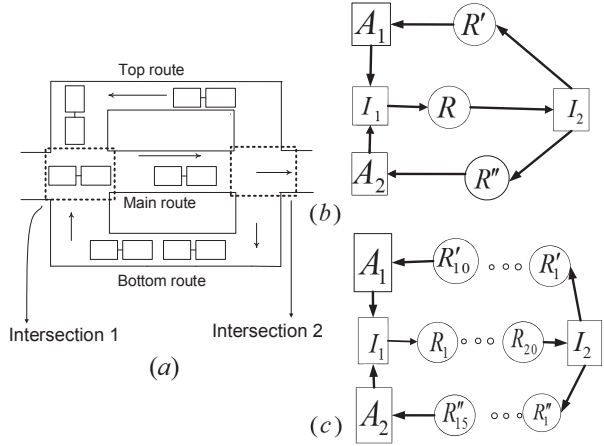

Fig. 1. (a) A traffic network consisting of two intersections and three routes. Spaces in each route get filled by arrival a car of a train from the previous space and become empty when a car passes to the next space. Arrows show the direction of train movements. (b) The PCN of traffic network example. $I_1$ is an input node and $I_2$ is an output node. $R$, $R'$, and $R''$ are parameterized nodes. The last two spaces of top (bottom) route are modeled by distinguished subprocesses $A_1$ ($A_2$). (c) An instance of the traffic network example where parameterized nodes $R$, $R'$, and $R''$ are replaced by parameterized segments with 20, 10 and 15 subprocesses respectively.

ranges over the set of natural numbers greater than two. Formally,

$$\mathcal{P} = \{\|_{i=1}^{M} P_i : \ M > 2\},$$

where $P_i = (X_i, \Sigma_i, \xi_i, x_i^0)$, with $X_1 = X_2 = ...$, and $M$ is the unspecified parameter. We are particularly interested in PDES with *linear* topology. PDES $\mathcal{P}$ has linear topology if for any member $\|_{i=1}^{M} P_i \in \mathcal{P}$, subprocess $P_i$, $1 < i < M$, has events shared only with both $P_{i-1}$ and $P_{i+1}$, and $P_1$ and $P_M$ respectively have events shared only with $P_2$ and $P_{M-1}$.

We assume all subprocesses have the same state set $X_s$ and instantiated from a template subprocess $P_n$ in the following manner. Let $P_n = (X_n, \Sigma_n, \xi_n, x_n^0)$, and assume all event symbols in $\Sigma_n$ have either $n$ or $n+1$ as indices. Define instance $P_i$ for any $i \in \mathbb{N}$, by replacing the index $n$ (respectively $n+1$) with $i$ (respectively $i+1$), and defining $\xi_i$ such that for all $x \in X_s$ and $\sigma_n \in \Sigma_n$ (respectively $\sigma_{n+1} \in \Sigma_n$), $\xi_i(x, \sigma_i) = \xi_n(x, \sigma_n)$ (respectively $\xi_i(x, \sigma_{i+1}) = \xi_n(x, \sigma_{n+1})$).

We set $\Sigma_i = \Sigma_{L_i} \cup \Sigma_{S_i}$; $\Sigma_{L_i}$ is the set of local events (events that are shared neither with $P_{i-1}$ nor with $P_{i+1}$) and $\Sigma_{S_i}$ is the set of shared event symbols. Local event alphabets are pairwise disjoint. Symbols in $\Sigma_{S_i}$ either have index $i$ or index $i+1$: shared events between subprocesses $P_{i-1}$ and $P_i$ have index $i$, while event shared between $P_i$ and $P_{i+1}$ have index $i+1$. In the example of Figure 1(a), each route has an arbitrary length and can be modeled as a linear PDES. Figure 2(b) depicts models of the $i^{th}$ and $(i+1)^{th}$ spaces of the main route.

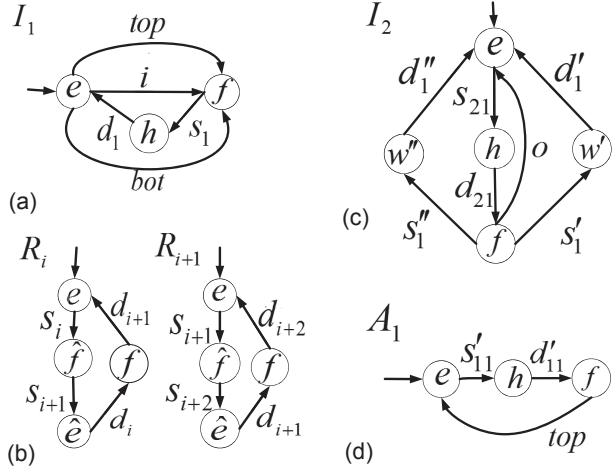

Fig. 2. The models of subprocesses of the traffic network example. (a) The distinguished subprocess that models intersection one. Event $i$ is a local event (the entrance of a train from the outside of the network). Shared events $top$ and $bot$ denote entrance of a train from the top and bottom routes. The intersection goes back to the empty state by departure of both cars of the train to the next space (string $s_1 d_1$, where $s_1$ and $d_1$ indicate departure of the first and second cars). (b) The models of the $i^{th}$ and $(i+1)^{th}$ spaces in the linear PDES representing the main route. The $i^{th}$ space gets filled by arrival of the first car (event $s_i$) and becomes empty by event $s_{i+1}$ and wait for the second car of the train, then the second car fills the space by event $d_i$ and it leaves by event $d_{i+1}$. (c) the model of intersection two. It gets filled by string $s_{21} d_{21}$. A train can exit the network by local event $o$. Alternatively, a train returns via the top (bottom) route by string $s_1' d_1'$ ($s_1'' d_1''$). (d) the model of $A_1$ (model of $A_2$ is similar) which indicates the last two spaces of the top route.

**Remark 1.** The assumption $M > 2$ is for efficient presentation of the results: our framework can be applied to networks with linear PDES segments with 1 or 2 subprocesses, however such networks may require different dependency graphs.

*C. Parameterized-chain networks*

A *PCN* is a strongly connected, finite, directed graph whose nodes are partitioned into *distinguished* nodes and *parameterized* nodes. The former, represented graphically as squares, will denote distinguished subprocesses, and the latter, represented as circles, will denote linear PDES that are subnetworks of the overall system. Distinguished nodes are finite-state subprocesses that can have a structure distinct from those of other subprocesses. Each parameterized node is the template finite-state subprocess for the linear PDES that the node denotes. All parameterized nodes have an in-degree and an out-degree of one. We assume that the state sets corresponding to subprocesses associated with different nodes are disjoint. We denote the (distinguished) nodes with in-degree larger than one *input nodes*, and the nodes with out-degree larger than one *output nodes*. We make the following structural assumptions on the PCN: it has a single input node, the input node is not an output node, and output nodes are not direct successors or direct predecessors of the input node. See Figure 1(b) for an example of PCN. In the running example of the traffic network (Figure 1), subprocesses $I_1$, $I_2$, $A_1$, and $A_2$ are distinguished subprocesses. $I_1$ is the input node and $I_2$ is an output node. $R$, $R'$, and $R''$ are parameterized nodes.

A PCN represents an infinite family of finite-state systems. Each member of a PCN family is represented by an *instance*, whose topology is inherited from that of the PCN: an instance is obtained from a PCN by 'expanding' each parameterized node into a finite, directed, linear subgraph with $M$ nodes for some particular value $M > 2$, where each node of the linear subgraph is a subprocess of the parameterized segment. The direction of the arcs in the linear subgraph agrees in the evident way with those of the unique arcs leading into and out of the corresponding parameterized node, so that the overall graph is, like the PCN from which it is derived, strongly connected.

Recall that all nodes of a PCN instance are subprocesses. When a parameterized node is expanded into a linear subgraph of $P_1, P_2, ..., P_M$, that leads to a distinguished node $D$, any occurrence in $D$ of an event $\sigma_{n+1}$ shared by $D$ and the template of the parameterized segment is replaced in $D$ by $\sigma_{M+1}$ (for example, see shared events $s_{21}$ and $d_{21}$ in the model of distinguished subprocess $I_2$, in an instance of a PCN depicted in Figure 2(c)).

Any two nodes of an instance that are connected by a single arc are called *neighbors*. Subprocesses have common shared events only if they are neighbors. We assume that each event symbol is at most shared between two subprocesses. The term input (output) subprocess in an instance refers to an input (output) node.

Each subgraph of a PCN instance corresponds to a generator obtained by the synchronous product of all subprocesses in that subgraph. We will not distinguish between a subgraph of an instance and its corresponding generator.

*D. Assumptions on a PCN*

Checking existence of deadlock in a parameterized network is undecidable even for the case of a parameterized network with ring topology [2]. In this paper, we consider parameterized-chain networks consisting of several parameterized segments as well as distinguished subprocesses with a more general topology. Thus, in order to characterize a tractable subproblem, we impose some restrictions on PCN. The following assumptions are expressed for any instance of a PCN; however the satisfaction of these assumptions for any instance implies their satisfaction in all instances (See Remark 3 of [2]).

First, we set mild assumptions on all subprocesses of all instances of the PCN (assumptions (1-3) below). Then we restrict the input subprocess by (4-5) and output subprocesses by (6).

Consider any instance of a PCN. Let $G_i$ and $G_{i+1}$ be two arbitrary neighboring subprocesses of this instance such that $G_{i+1}$ is a direct successor of $G_i$. For $k = i, i+1$, let $G_k = (X_k, \Sigma_k, \xi_k, x_k^0)$. We assume the following:

$$(\forall x_i, x_i' \in X_i)(\exists t \in \Sigma_i^*)[x_i' \in \xi_i(x_i, t)], \tag{1}$$

$$(\forall \sigma_i \in \Sigma_i \cap \Sigma_{i+1})[|\chi_i(\sigma_i)| = 1], \tag{2}$$

$$(x_i^0, x_{i+1}^0) \in \mathcal{V}_{i+1} \tag{3}$$

where $\mathcal{V}_{i+1}$ is a weak invariant simulation of $G_{i+1}$ by $G_i$ w.r.t. $\Sigma_i \cap \Sigma_{i+1}$.

Assumption (1) is a condition on the structure of individual subprocess $G_i$, while assumptions (2) and (3) restrict the way subprocesses interact. Assumption (1) states that the transition graph of each subprocess is strongly connected. This assumption often holds in nonterminating subprocesses: in the absence of synchronization with other subprocesses, this assumption rules out states that could become permanently inaccessible as the subprocess evolves.

By (2), each shared event in the subset $\Sigma_i \cap \Sigma_{i+1}$ has exactly one companion state in subprocess $G_i$. In other words, interactions between $G_i$ and $G_{i+1}$ via a specific shared event in $\Sigma_i \cap \Sigma_{i+1}$ can occur only if $G_i$ is in that specific state. If (2) is not satisfied, suitable enrichment of the event alphabet would make it hold (by distinguishing occurrences of the same event that can occur in distinct states), but this alphabet enrichment could make the remaining assumption (3) stronger.

Assumption (3) states that $G_i$ weakly invariantly simulates $G_{i+1}$ with respect to $\Sigma_i \cap \Sigma_{i+1}$. This assumption implies a sense of directionality between neighboring subprocesses of the network. It expresses that $G_i$ can eventually execute any event shared with $G_{i+1}$, if interaction with the rest of the network is ignored. Violation of assumption (3) means that even if the interaction of $G_i$ and $G_{i+1}$ with the rest of the network is ignored, $G_i$ may never be able to provide some of the resources needed by $G_{i+1}$. This might indicate a 'design flaw' in network architecture that can easily be identified by calculation of the synchronous product of $G_i$ and $G_{i+1}$. Assumption (3) usually holds in networks that contain 'directional' parameterized segments; for example, in many manufacturing plants, workpieces normally move in a default direction and a subprocess can always expect eventually to receive a workpiece from its direct predecessor neighbor. In the traffic network example of Figure 1, where routes are modeled as linear parameterized segments, this assumption implies a space in a route eventually receives train cars from the previous space (see the modeling of Figure 2(b)).

To make the analysis tractable, we now restrict the structure of the input and output subprocesses. Consider an arbitrary instance of a PCN. Let $G_1$ be the unique input subprocess of an instance, and $G_2$ be its direct successor subprocess, and $G_N$ be any of its direct predecessors in the instance. Let $G_k = (X_k, \Sigma_k, \xi_k, x_k^0)$, $k = 1, 2, N$. We assume

$$(\forall \alpha \in \Sigma_N \cap \Sigma_1)(\forall \beta \in \Sigma_1 \cap \Sigma_2)[\chi_1(\alpha) \cap \chi_1(\beta) = \emptyset], \tag{4}$$

$$(\forall (x_1, x_2) \in R)[x_1 \mathcal{W} x_2], \tag{5}$$

where $R$ is the state set of synchronous product $G_1 \| G_2$ and $\mathcal{W}$ is a weak invariant simulation of $G_2$ by $G_1$ w.r.t. shared events of $G_1$. Assumption (4) expresses that for any state of $G_1$ in which an event shared with $G_2$ is enabled, there is no event shared with $G_N$ enabled from that state, and vice versa. In the traffic network example of Figure 1, this assumption implies that when intersection one (the input subprocess) is in state $f$, from which the shared event of a train exiting from the intersection (event $d_1$) is enabled, there is no event shared with the top and bottom routes that can occur. This means that a train from these routes cannot enter the intersection when it is full (in state $f$).

Assumption (5) expresses that all the state pairs in the synchronous product of $G_1$ and $G_2$ are in relation $\mathcal{W}$. This means that from any reachable global state, if $G_2$ is in a state in which a shared event with $G_1$ is defined, $G_1$ can always reach the companion state of that shared event without executing any other shared event. In other words, the input subprocess $G_1$ acts a source node: regardless of the states of the rest of the network, $G_1$ can always provide resources requested by $G_2$. Note that this assumption is stronger than (3) and further reinforces the directionality of the network. Although this assumption on the unique input node is relatively strong, it is a natural assumption for some networks. For example in a manufacturing pipeline, this assumption implies an inexhaustible source of workpieces entering the pipeline. In the traffic network example of Figure 1, it implies possible entrance of a train into the traffic network at any time. This assumption is used to establish reachability of the generalized circular waits that we compute below. If it is relaxed, the method may compute some generalized circular waits that are in fact unreachable. This may represent a useful compromise for purposes of control synthesis, where at worst it will lead to a control policy that is more restrictive than strictly necessary.

Let $G_j$ be an arbitrary output subprocess in an instance of a PCN, $G_{j+1}$ be any of its direct successor subprocesses and $G_{j-1}$ be its predecessor subprocess. For any such $G_{j-1}$, $G_j$ and $G_{j+1}$, we assume

$$(x_j^0, x_{j+1}^0) \in \mathcal{Q}_{j+1}, \tag{6}$$

where $\mathcal{Q}_{j+1}$ is a weak invariant simulation of $G_{j+1}$ by $G_j$ w.r.t. $\Sigma_{S_j} \setminus \Sigma_{j-1}$.

Assumption (6) determines how output subprocess $G_j$ interacts with its direct successor subprocesses in any instance of the PCN. This assumption expresses that output subprocess $G_j$, from its initial state, can reach companion states of events shared between $G_j$ and $G_{j+1}$ via a string that contains no event shared with its other direct successor subprocesses (different from $G_{j+1}$). However, the simulation relation $\mathcal{Q}_{j+1}$ need not hold after $G_j$ executes an event shared with other direct successor subprocesses. In other words, as long as $G_j$ executes no event shared with its direct successor subprocesses other than $G_{j+1}$, execution of events shared between $G_j$ and $G_{j+1}$ is not blocked by its other direct successor subprocesses. However, subprocess $G_j$ may execute a shared event with the rest of the network at any time, after which the simulation relation need not hold. In the traffic network example of Figure 1, $R_1'$ and $R_1''$ are the direct successors of output subprocess $I_2$. In this example assumption (6) means that the first space of each route can expect to receive the second car of the train after receiving the first one.

## IV. THE DEADLOCK ANALYSIS

In the present section we characterize a generalized version of reachable circular waits among subprocesses of any instance of a PCN. Specifically, we define for any PCN a *dependency graph* on states of subprocesses, and define 'full, consistent' subgraphs of the dependency graph as a tool for detection of partial and total deadlocks of PCN instances. Partial and total deadlocks in an instance of a PCN are formally defined below.

**Definition 3.** Let $X'$ be the state set of a subgraph of a PCN instance (that is, the Cartesian product of the state sets of these subprocesses); then $x \in X'$ is a *partial deadlock* of that PCN instance if under synchronization with the rest of the network, subprocesses of the subgraph can reach state $x$, but no transition is possible from that state. A partial deadlock is a *total deadlock* if the subgraph is the entire instance. An instance of a PCN is *deadlock-free* if it has no total deadlock.

### A. Cycles and isolated cycles

In order to locate reachable circular waits among the subprocesses, we initially focus on the individual 'cycles' of instances of a PCN. For the purpose of our analysis, we disable certain transitions of input and output subprocesses to yield a subgraph with ring structure. The next operation will be used to restrict the transitions of subprocesses.

**Definition 4.** For a given generator $G_i = (X_i, \Sigma_i, \xi_i, x_i^0)$, $G_i(\xrightarrow{\Delta_i})$ is the restriction of the generator to a transition function $\hat{\xi} : X_i \times \Delta_i \to 2^{X_i}$ and is formed by erasing transitions with events that belong to the set $\Sigma_i \setminus \Delta_i$, and unreachable states. Formally, $G_i(\xrightarrow{\Delta_i}) = (\hat{X}_i, \Sigma_i, \hat{\xi}_i, x_i^0)$ and, for all $x_i \in X_i$ and $\sigma \in \Sigma_i$,

$$\hat{\xi}_i(x_i, \sigma) = \begin{cases} \xi_i(x_i, \sigma), & \text{if } \sigma \in \Delta_i; \\ \emptyset, & \text{if } \sigma \in \Sigma_i \setminus \Delta_i; \end{cases}$$

and $\hat{X}_i \subseteq X_i$ is the set of all $\hat{x}_i \in X_i$, such that there exists $l \in \Sigma_i^*$ for which $\hat{x}_i \in \hat{\xi}_i(x_i^0, l)$.

Note that the above operation does not alter the alphabet of $G_i$; it merely prevents the occurrence of any events in $\Sigma_i \setminus \Delta_i$ by altering the transition function of the generator.

Next, we define a *cycle* and an *isolated cycle* of a PCN. These notions are defined with reference to an instance (not the PCN itself). When we refer to a cycle or isolated cycle with $N$ subprocesses, terms $i+j$ and $i-j$ are calculated using *modulo-N* arithmetic over the complete residue system $\{1, 2, ..., N\}$.

**Definition 5.** A *cycle* $G^N = \|_{i=1}^{N} G_i$ of an instance of a PCN is the synchronous product of $N$ subprocesses of a simple circuit in an instance, with the respective distinct subprocesses relabeled from $G_1$ to $G_N$ in the direction of the arcs, starting with the input subprocess. (Note that any simple circuit in an instance must include the unique input node.) Let $G_i = (X_i, \Sigma_i, \xi_i, x_i^0)$, $1 \leq i \leq N$, and $\Delta_i = (\Sigma_{i-1} \cap \Sigma_i) \cup (\Sigma_i \cap \Sigma_{i+1}) \cup \Sigma_{L_i}$. Then the *isolated cycle* is $\hat{G}^N = \|\hat{G}_i = (\hat{X}, \Sigma, \hat{\xi}, x^0)$, where $\hat{G}_i = G_i(\xrightarrow{\Delta_i})$. States of cycles or isolated cycles with $N$ subprocesses take the form of N-tuples $x = (x_1, x_2, ..., x_N)$, where $x_i$ is the state of the $i^{th}$ subprocess $G_i$.

**Remark 2.** An isolated cycle $\hat{G}^N$ is a ring network formed by restriction of all subprocesses of the cycle $G^N$ to the transitions that are not shared with subprocesses outside of $G^N$. Note that in $G^N$, the only subprocesses that have events shared with subprocesses outside of $G^N$ are the input subprocess $G_1$ and output subprocesses $G_j$, $j \in J$, where $J$ is the index set of output subprocesses of $G^N$.

## B. Forward dependency property

Here we define a *forward dependency* property based on synchronous products of neighboring subprocesses in an isolated cycle of the network. This property aims to characterize the occurrence of a circular wait. By Lemma 1 given in the appendix, in an isolated cycle all the events of a subprocess shared with the neighbor of 'lower' index can eventually be executed. Therefore the only shared events that may be blocked in an isolated cycle are those shared with the neighbor of 'larger' index. A state pair $(x_{i-1}, x_i)$ in a synchronous product of two neighboring subprocesses $G_{i-1}$ and $G_i$ is forward-dependent if the only event enabled in state $(x_{i-1}, x_i)$ in the synchronous product is an event shared with $G_{i+1}$.

**Definition 6.** Consider cycle $G^N = \|G_i, 1 \leq i \leq N$ of an instance of a PCN and isolated cycle $\hat{G}^N = (\hat{X}, \Sigma, \hat{\xi}, x^0)$. For any $i$, $1 \leq i \leq N$, let $\hat{R}_i$ be the state set of the synchronous product $\hat{G}_{i-1} \| \hat{G}_i = (\hat{R}_i, \Sigma_{i-1} \cup \Sigma_i, \delta_i, (x_{i-1}^0, x_i^0))$. A state pair $(x_{i-1}, x_i) \in \hat{R}_i$ is *forward-*

*dependent* if

$$(\forall \sigma_i \in \Sigma_{i-1} \cup \Sigma_i)[(\delta_i((x_{i-1},x_i), \sigma_i) \neq \emptyset)$$
$$\Rightarrow (\chi_{i+1}(\sigma_i) \neq \emptyset)]. \tag{7}$$

A state $x \in X_1 \times X_2 \times ... \times X_N$ is forward-dependent if for all $i$, $1 \leq i \leq N$, $(x_{i-1}, x_i) \in \hat{R}_i$ and $(x_{i-1}, x_i)$ satisfies (7). We denote by $X_d \subseteq X_1 \times X_2 \times ... \times X_N$ the set of all forward-dependent states of cycle $G^N$.

If a state pair $(x_{i-1}, x_i) \in \hat{R}_i$ satisfies (7), it means that the only transitions available from this pair in the synchronous product $\hat{G}_{i-1} \| \hat{G}_i$ are shared with the neighbor of 'larger' index in the isolated cycle, $\hat{G}_{i+1}$. For a reachable state $x$ in $\hat{G}^N$, if property (7) holds for all $i$, then all the subprocesses of $\hat{G}^N$ are waiting for execution of an event shared with their respective immediate neighbors with larger index. Note that for such a state $x$, there can be events shared with subprocesses outside of cycle $G^N$ enabled from $x_1$ or $x_j$, $j \in J$, where $J$ is the index set of output subprocesses of $G^N$. Execution of these shared events may break the circular wait within the cycle. Therefore, the existence of a circular wait in an isolated cycle of the network need not cause a partial deadlock. We introduce the *dependency graph* below to identify generalized circular waits among multiple cycles of the network which cause a partial deadlock.

## C. The dependency graph

Define the binary relation *Depend* to be the set of all forward-dependent state pairs in any isolated cycle in any instance. The *dependency graph* is based on this relation: its nodes are exactly those states that belong to any pair in *Depend*; its arcs are precisely the elements of *Depend*.

Note that the relation *Depend* can be computed by considering a single, arbitrary instance. The state sets of all subprocesses in any parameterized segment are the same and these subprocesses are isomorphic. Therefore the forward-dependent pairs in any two neighboring subprocesses (except the last two) in a parameterized segment are the same. Consequently, any instance includes all of the pertinent local structure that appears in any cycle of any other instance. Hence the dependency graph can be constructed based on an instance of minimal size (where each parameterized node is replaced by a linear PDES with three subprocesses).

The computation of the set *Depend* is as follows. Consider all isolated cycles of the instance of PCN with minimal size. For each isolated cycle, consider all sequences of three consecutive subprocesses. For each of them, compute a synchronous product of the first two subprocesses, and check forward dependency by comparison of the result to the third subprocess. The computational complexity of this is polynomial-time in the maximum subprocess alphabet size, the subprocess state-set cardinality, the sum of out-degree of output nodes, and the number of nodes of PCN.

The next definition states the *consistency* property of a subgraph of the dependency graph and explains how a consistent subgraph of the dependency graph *represents* a set of states of subgraphs of instances of the PCN.

**Definition 7.** A subgraph $\bar{\mathcal{D}}$ of the dependency graph $\mathcal{D}$, is *consistent* if it is strongly connected and contains a state of the input node and does not contain more than one state of any distinguished subprocess.

A consistent subgraph $\bar{\mathcal{D}}$ is perhaps best thought of as representing a regular set of 'putative' states of strongly connected subgraphs of instances of the PCN. (We say 'putative' because it is not clear a priori that these 'states' are reachable – their reachability is established below).

Indeed a consistent subgraph of the dependency graph uniquely determines a state of the input node and of other distinguished subprocesses. Within such a subgraph $\bar{\mathcal{D}}$ the existence of loops consisting of nodes that are states of subprocesses belonging to the same linear parameterized segment reflects the arbitrary length of instances of that segment. A consistent subgraph thus determines states of distinguished subprocesses and regular sets of possible states of linear parameterized segments linking those distinguished subprocesses.

The consistent subgraph $\bar{\mathcal{D}}$ is said to *represent* all putative states of strongly connected subgraphs of the instances of PCN in which:

(a) The states of distinguished subprocesses are only those determined by $\bar{\mathcal{D}}$; and

(b) The states of instances of parameterized segments consist exactly of one member of each of the regular sets determined by $\bar{\mathcal{D}}$.

Consider Figure 3 (a), showing the dependency graph of the traffic network example. The consistency property contains two main conditions. First, the subgraph must be strongly connected and must include the input node. For example, in the dependency graph of Figure 3, the loop between nodes $f_R$ and $\hat{f}_R$ is not a consistent subgraph because it does not include the input node. Accordingly, it does not correspond to a circular wait. Indeed it represents only parameterized segment $R$ in which all subprocesses are alternating in state $f$ and $\hat{f}$ (note that the state sets of all subprocesses of the linear parameterized segment are the same). The second condition of consistency is that the dependency graph does not include two states for the same distinguished subprocess (obviously a subprocess cannot be in two states simultaneously). In the dependency graph of Figure 3(a), a consistent subgraph cannot contain both nodes $w'_{I_2}$ and $w''_{I_2}$ or both $f_{I_1}$ and $h_{I_1}$. This dependency graph contains four consistent subgraphs. The two such subgraphs that present the states of the top loop of the network are depicted in Figure 3 parts (b) and (c) (the two that present the bottom cycle are similar owing to network symmetry). Each of these subgraphs represents the states of a set of instances of the top loop of the traffic network example. For example, subgraph 3(b) represents the following state set: $I_1$ is in state $f$, the main route comprises an arbitrary number $M$ of spaces, and the corresponding subprocesses $R_i$, $1 \leq i \leq M$, are in state $f$ for odd values of $i$ and in state $\hat{f}$ for even values of $i$. Subprocess $I_2$ is in state $w'$, the top route consists of $M'$ spaces and corresponding subprocesses $R'_i$, $1 \leq i \leq M'$, are in state $\hat{f}$ for odd values of $i$ and in state $f$ for even values of $i$. Finally $A_1$ is in state $f$.

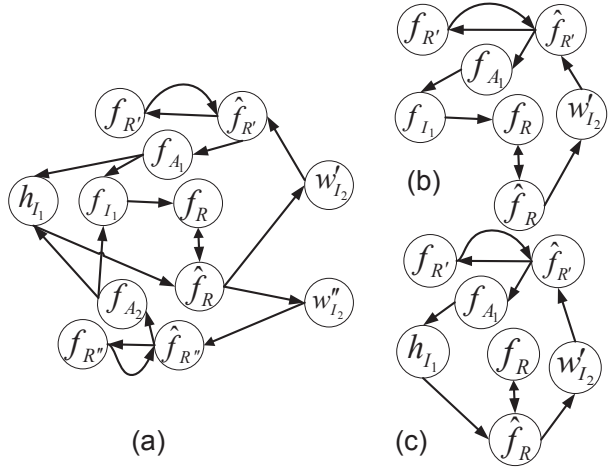

Fig. 3. (a) The dependency graph of traffic network example. (b),(c) Two of full, consistent subgraphs of the dependency graph of traffic network example.

**Remark 3.** In the consistent subgraphs of Figure 3 parts (b) and (c), $(w'_{I_2}, \hat{f}_{R'})$ is the only outgoing arc from node $w'_{I_2}$. Therefore in all states represented by this subgraph, the first subprocess of parameterized segment $R'$ is in state $\hat{f}$. Then $(\hat{f}_{R'}, f_{R'})$ is the only outgoing arc from $\hat{f}_{R'}$ to the state of a subprocess in $R'_1$. Therefore in all states presented by these subgraphs, the second subprocess of $R'$ is in state $f$. Using this argument, we conclude that subprocesses $R'_i$ are in state $\hat{f}$, for odd values of $i$ and in state $f$ for even values of $i$. Since $(\hat{f}_{R'}, f_{A_1})$ is the only arc from states of the linear PDES $R'$ to $A_1$, the last subprocess of $R'$ in all states represented by this subgraph is in state $\hat{f}$. This means that $R'$ has an odd number of subprocesses. Hence this subgraph represents no state of instances of the PCN with even numbers of subprocesses in linear PDES $R'$. Similarly, all other consistent subgraphs of the dependency graph represent sets of states of instances of the PCN with odd numbers of subprocesses of linear PDES segment $R''$.

## D. Deadlock detection

As mentioned earlier, forward-dependent states of isolated cycles that form the dependency graph need not represent partial deadlocks of an instance of a PCN. To establish a relationship between the dependency graph and reachable total and partial deadlocks, we now define the *full* subgraphs of the dependency graph. This property deals with the issue of output subprocesses in deadlock analysis: a state of an output subprocess may have events shared with different direct successor subprocesses. In order to prevent execution of these shared events, states of all direct successors must be included in a suitably generalized circular wait. Therefore a corresponding subgraph of the dependency graph has to include branches that correspond to each of these direct successors.

**Definition 8.** A subgraph of a dependency graph is *full* if, for any state $x_j$ of any output subprocess $G_j$, and any direct successor $G_{j+1}$ of $G_j$, if an event shared with $G_{j+1}$ is enabled from $x_j$ in $G_j$, then the subgraph contains exactly one arc $(x_j, x_{j+1})$ where $x_{j+1}$ is a state of $G_{j+1}$.

All consistent subgraphs of the dependency graph of the traffic network example are also full. Figure 3 parts (b) and (c) show two of these full, consistent subgraphs. The relationship between full, consistent subgraphs of the dependency graph and deadlocks in PCN is described below.

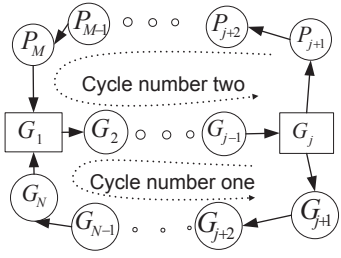

Fig. 4. A GPDES instance containing a single output subprocess.

## V. Main results

To aid readability and avoid cumbersome notation, we first consider the particular network structure of Figure 4 and carry out the deadlock analysis (Theorem 1). We then use the results of this analysis for development of a deadlock analysis method for a general GPDES (Theorem 2).

The deadlock analysis involves the following question: is a forward-dependent state represented by a dependency graph in fact a reachable deadlocked state? The next theorem provides a response for the case of the network structure of Figure 4.

**Theorem 1.** Consider a GPDES satisfying (1-6) with the structure of Figure 4. Let $\mathscr{D}$ be the dependency graph of this GPDES. Denote cycle number one as $G^N = (X, \Sigma, \xi, x^0)$. In an instance of this GPDES let $x$ and $y$ respectively be the states of cycles one and two in Figure 4.

(a) Assume there is no event enabled from $x_j$ in output subprocess $G_j$ that is shared with subprocesses outside of cycle number one. State $x$ is a partial deadlock if and only if it is represented by a full, consistent subgraph of $\mathscr{D}$.

(b) A state of an instance of the GPDES is a total deadlock if it is represented by a full, consistent subgraph of $\mathscr{D}$.

*Proof. Part (a)*: (*If*) Assume $x$ is represented by a full, consistent subgraph of $\mathscr{D}$, but is not a partial deadlocked state of the GPDES instance. According to Proposition 5(c), $x$ is a forward-dependent state. Therefore by Lemma 2, state $x$ is reachable in $\hat{G}^N$, and therefore is reachable within the global network. By assumption, state $x$ is not deadlocked; therefore for some event $\beta \in \Sigma$, $\xi(x, \beta) \neq \emptyset$. By the structure of the network, only $G_1$ and $G_j$ have events shared with subprocesses outside of cycle number one. But by assumption, there is no event enabled from $x_j$ in $G_j$ that is shared with subprocesses outside of $G^N$. By (5) and the definition of forward-dependent states, there is an event shared with $G_2$ enabled from $x_1$ in $G_1$. Therefore by (4), there is no event enabled from $x_1$ that is shared with $P_M$. This means that $\beta$ is an event defined in one of the subprocesses of isolated cycle $\hat{G}^N$. Since $x$ is a forward-dependent state, by (7) $\beta$ must be a shared event. Let $i$, $1 \leq i \leq N$ be such that $\beta \in \Sigma_{i-1} \cap \Sigma_i$. Given that $\xi(x, \beta) \neq \emptyset$, if $\delta_i$ is the transition function of $\hat{G}_{i-1} \| \hat{G}_i$, we have $\delta_i((x_{i-1}, x_i), \beta) \neq \emptyset$. But by (7), $\chi_{i+1}(\beta) \neq \emptyset$. This means that a $\beta$ transition is defined both in $\hat{G}_{i+1}$ and in $\hat{G}_{i-1}$. This contradicts the network assumption that only neighboring subprocesses have events shared between them. Therefore state $x$ is a reachable deadlocked state of cycle number one and partial deadlocked state of the GPDES instance.

(*Only if*) Consider an arbitrary reachable partial deadlocked state $x \in X$.

We first show that state $x$ belongs to the subset $X_d$ of Definition 6. Because $x$ is a reachable state, for all $i$, $1 \leq i \leq N$,

$$(x_{i-1}, x_i) \in R_i, \tag{8}$$

where $R_i$ is the state set of the synchronous product $G_{i-1} \| G_i = (R_i, \Sigma_{i-1} \cup \Sigma_i, \delta_i, (x_{i-1}^0, x_i^0))$.

We next show that for all $i$, any events enabled from $x_i$ in $G_i$ must belong to $\Sigma_i \cap \Sigma_{i+1}$. To do so, we prove that $(x_j, x_{j+1}) \in \mathcal{Q}_{j+1}$, where $\mathcal{Q}_{j+1}$ is a weak invariant simulation w.r.t $\Sigma_{S_j} \setminus \Sigma_{j-1}$. According to (1), there must exist an event enabled from $x_j$. By assumption there is no event enabled from $x_j$ that is local or shared with $P_{j+1}$. By Proposition 1, $(x_{j-1}, x_j) \in \mathcal{V}_j$, where $\mathcal{V}_j$ is a weak invariant simulation of $G_j$ by $G_{j-1}$ w.r.t. $\Sigma_{j-1} \cap \Sigma_j$. So there exists an event $\sigma_{j-1} \in \Sigma_{j-1} \cap \Sigma_j$ such that $\xi_j(x_j, \sigma_{j-1}) \neq \emptyset$, then by definition of weak invariant simulation, there exists a string $l_{j-1} \in (\Sigma_{j-1} \setminus \Sigma_j)^*$ such that $\chi_{j-1}(\sigma_{j-1}) \in \xi_{j-1}(x_{j-1}, l_{j-1})$. If $l_{j-1}$ has no events shared with $G_{j-2}$, then $\sigma_{j-1}$ can be executed, a contradiction. For the case $l_{j-1}$ contains shared events with $G_{j-2}$, by Lemma 1 (because $G_{j-1} = \hat{G}_{j-1}$), these shared events can be executed. Therefore the only events defined from $x_j$ are shared with $G_{j+1}$. But this means that $(x_j, x_{j+1})$ cannot belong to any weak invariant simulation w.r.t $\Sigma_{S_j} \setminus \Sigma_{j-1}$. According to Proposition 4, we must have

$$(x_j, x_{j+1}) \in \mathcal{Q}_{j+1}, \tag{9}$$

where $\mathcal{Q}_{j+1}$ is a weak invariant simulation w.r.t $\Sigma_{S_j} \setminus \Sigma_{j-1}$. Now set state $x$ as the initial state of $G^N$ to form a new cycle $G'^N$ and its isolated version $\hat{G}'^N$. By Lemma 1, if the event enabled from $x_i$, $1 \leq i \leq N$, is shared with $\hat{G}'_{i-1}$, it can be executed, a contradiction. Because $G'_1$ and $G'_j$ are respectively the only input and output subprocesses, for all $i \notin \{1, j\}$ $\hat{G}'_{i-1} = G'_{i-1}$ (see Remark 2). This means that for $i \notin \{1, j\}$ any events enabled in $x_i$ must belong to $\Sigma_i \cap \Sigma_{i+1}$. We have already shown that the only events enabled from $x_j$ in $G_j$ are shared with $G_{j+1}$. For $i = 1$, by Proposition 2(b), $(x_1 \hat{\mathcal{W}} x_2)$, where $\hat{\mathcal{W}}$ is a weak invariant simulation of $\hat{G}_2$ by $\hat{G}_1$ w.r.t. all shared events of $G_1$. This means that there must be an event defined from $x_1$ in $\hat{G}'_1$. By assumption, no local event is enabled from $x_1$ in $G_1$. So, in order for weak invariant simulation $\hat{\mathcal{W}}$ to hold, there must be an event shared with $G_2$ enabled from $x_1$ in $G_1$. By (4) there is no other event enabled from $x_1$ in $G_1$. Therefore the only event enabled from $x_1$ is shared with $G_2$; so for all $i$, $(x_{i-1}, x_i)$ satisfy (7).

Now it suffices to show that for all $i$, $(x_{i-1}, x_i) \in \hat{R}_i$, where $\hat{R}_i$ is the state set of the synchronous product $\hat{G}_{i-1} \| \hat{G}_i$. Because $G_1$ and $G_j$ are the only subprocesses that have shared events with subprocesses outside of $G^N$, for $i \neq 1, j$, subprocesses $G_i$ and $\hat{G}_i$ are the same. Therefore we only need to show $(x_{i-1}, x_i) \in \hat{R}_i$ for $i = 1, 2, j, j+1$. Note that by (8), (3) and Proposition 1, $(x_{i-1}, x_i) \in \mathcal{V}_i$.

For the case where $i = j+1$ or $i = 2$, by (1), there exists $l_i \in (\Sigma_i \setminus \Sigma_{i-1})^*$ such that for some $\sigma \in \Sigma_{i-1} \cap \Sigma_i$, $\xi_i(x_i, l_i \sigma) \neq \emptyset$. Therefore, because events enabled from $x_{i-1}$ are in $\Sigma_{i-1} \cap \Sigma_i$, by (9) for the case $i = j+1$, and by (5) for the case $i = 2$, we must have $\xi_{i-1}(x_{i-1}, \sigma) \neq \emptyset$. Let $s_i$ be the string labeling any path from $x_i^0$ to $x_i$. Consider string $s_i l_i$. Since $\xi_i(x_i^0, s_i l_i) \neq \emptyset$, by (6), there must exist a string $s_{i-1} \in \Sigma_{i-1}^*$ such that $\xi_{i-1}(x_{i-1}^0, s_{i-1} \sigma) \neq \emptyset$ and $P_{\Sigma_{S_{i-1}} \setminus \Sigma_{i-2}}(s_{i-1}) = P_{\Sigma_{S_{i-1}} \setminus \Sigma_{i-2}}(s_i l_i)$. But $l_i \in (\Sigma_i \setminus \Sigma_{i-1})^*$; therefore

$$P_{\Sigma_{S_{i-1}} \setminus \Sigma_{i-2}}(s_{i-1}) = P_{\Sigma_{S_{i-1}} \setminus \Sigma_{i-2}}(s_i). \tag{10}$$

Since $s_i \in \Sigma_i^*$, $s_{i-1}$ contains no event shared with $P_i$. By (2), the companion state of $\sigma$ in $G_{i-1}$ is unique; therefore $x_{i-1}$ is reachable in $\hat{G}_{i-1}$ via $s_{i-1}$. By (10), $P_{\Sigma_{i-1} \cap \Sigma_i}(s_{i-1}) = P_{\Sigma_{i-1} \cap \Sigma_i}(s_i)$. Therefore $(x_{i-1}, x_i) \in \hat{R}_i$.

Now for $i = j$ or $i = 1$, by (1) and (2), there exist $\sigma' \in \Sigma_{i-1} \cap \Sigma_i$ and $l_i \in \Sigma_i \setminus \Sigma_{i-1}$ s.t. $\xi_i(x_i, l_i \sigma') \neq \emptyset$. Because $(x_{i_1}, x_i) \in \mathcal{V}_i$, and because all events enabled from $x_i - 1$ belong to $\Sigma_{i-1} \cap \Sigma_i$, we must have $\xi_{i-1}(x_{i-1}, \sigma') \neq \emptyset$. Moreover, because $(x_{i-1}^0, x_i)$ belongs to a similar weak invariant simulation, there exists $s_{i-1}$ s.t. $\xi_{i-1}(x_{i-1}^0, s_{i-1} \sigma') \neq \emptyset$, with the same projection as $s_i$ onto $\Sigma_{i-1} \cap \Sigma_i$. Hence $(x_{i-1}, x_i) \in \hat{R}_i$. Therefore for all $i$, $1 \leq i \leq N$,

$$(x_{i-1}, x_i) \in \hat{R}_i, \tag{11}$$

*Part (b)*: We assume that $x$ and $y$ are components of a (putative) state represented by a full, consistent subgraph of $\mathscr{D}$, and show that they are components of a total deadlock ($x_1$ to $x_j$ are the same as $y_1$ to $y_j$). By Proposition 5(c) states $x$ and $y$ are forward-dependent states of cycles numbers one and two respectively; by Proposition 5(b) some events $\alpha \in \Sigma_{j+1}$ and $\beta \in \Sigma_{P_{j+1}}$ are enabled from $x_j$ in $G_j$. Furthermore, by the forward-dependency property no local events or events in $\Sigma_{j-1}$ are enabled from $x_j$. For simplicity, assume that $\alpha$ and $\beta$ are the only events enabled from $x_j$ in $G_j$. Event $\beta$ is shared with $P_{j+1}$, and $P_{j+1}$ to $P_M$ are in states $y_j$ to $y_M$. Since $y$ is a forward-dependent state, the only events enabled from $y_k$, $1 < k < j$ and $j+1 \leq k \leq M$, are shared with the respective neighbors of 'larger' index. This means that the only shared event enabled from $y_M$ is shared with $G_1$. By forward dependency and (4), the only shared event enabled from $y_1$ is shared with $G_2$. This forms a circular wait. Therefore no event other than $\alpha$ can occur from state $y$. By the same argument, no event other than $\beta$ can occur from state $x$. Therefore, if the putative state with components $x$ and $y$ is reachable, then it is a total deadlock.

Now we show that states $x$ and $y$ are indeed simultaneously reachable. By the fact that $x$ is a forward-dependent state, for all $i$, we have $(x_{i-1}, x_i) \in \hat{R}_i$, where $\hat{R}_i$ is the state set of the synchronous product $\hat{G}_{i-1} \| \hat{G}_i$. Therefore by Proposition 3, we have $(x_{i-1}, x_i) \in \mathcal{V}'_i$, where $\mathcal{V}'_i$ is a weak invariant simulation of $\hat{G}_i$ by $\hat{G}_{i-1}$ w.r.t. $\Sigma_{i-1} \cap \Sigma_i$. By Proposition 2(b), (5) also holds for this isolated cycle. Assumptions (1,2,3,4) are not affected by the isolation operation. By Proposition 5(c) and Lemma 2 state $x$ of cycle one is reachable in the isolated cycle. Now consider the pair $(x_j, y_{j+1})$.

By the same argument we have $(x_j, y_{j+1}) \in \mathcal{V}'_{j+1}$, where $\mathcal{V}'_{j+1}$ is a weak invariant simulation of $\hat{P}_{j+1}$ by $\hat{G}_j$ w.r.t. the shared events of $G_j$ and $P_{j+1}$. Set state $x$ as the new initial states for cycle number one and isolated cycle number two with these initial states. By the same reasoning as above, state $y$ is also reachable in this isolated cycle. Since $x_1$ to $x_j$ are the same as $y_1$ to $y_j$, states $x$ and $y$ are simultaneously reachable.

□

Theorem 1 relates to a network with the particular structure of Figure 4. For states $x$ such that no event enabled from state $x_j$ in $G_j$ that is shared with subprocesses outside of $G^N$, part (a) of the theorem provides a necessary and sufficient deadlock condition for partial deadlock. Part (b) gives a sufficient condition for total deadlock of the network.

The following theorem considers PCN with a generalized topology, and establishes that (a) any full, consistent subgraph of the dependency graph represents a partial deadlock; and (b) a necessary condition for occurrence of a total deadlock is existence of a full, consistent subgraph of the dependency graph. Note that partial and total deadlocks are reachable by definition.

**Theorem 2.** Consider a PCN $\mathcal{G}$ satisfying (1-6).
(a) Let $\mathcal{S}$ be a full, consistent subgraph of the dependency graph of $\mathcal{G}$. Then every state represented by $\mathcal{S}$ is a partial deadlock of an instance of $\mathcal{G}$.
(b) An instance of $\mathcal{G}$ has a total deadlock only if a state of a subgraph of the instance is represented by a full, consistent subgraph of the dependency graph of $\mathcal{G}$.

*Proof. (Part (a))* We first show that such a state (if reachable) is a partial deadlock, then we show its reachability. The proof of this part is similar to that of Theorem 1(b). Let $x$ be a state represented by $\mathcal{S}$. Since $\mathcal{S}$ is consistent, it is strongly connected. Therefore any subgraph of an instance of which $x$ is a state must also be strongly connected. Let $x_j$ be the (unique) state of an output subprocess $G_j$ represented by $\mathcal{S}$. By fullness the instance subgraph must include any direct successor $G_{j+1}$ of $G_j$ (provided that an event shared with $G_{j+1}$ is enabled from $x_j$ in $G_j$).

Because $\mathcal{S}$ is a subgraph of the dependency graph, no subprocess of the instance subgraph can execute until one of its direct successors does. By strong connectedness of the instance subgraph, this can never happen. Therefore $x$ is in a generalized circular wait and consequently deadlocked.

To show reachability of states represented by $\mathcal{S}$, assume that $x$ is such a state. We show that $x$ is reachable in $\mathcal{G}$. The proof is by induction on the structure of this subgraph. Since $\mathcal{S}$ is consistent, it contains a cycle that includes the input node. Add to this cycle all arcs $(u, v)$ such that $u$ and $v$ are states of parameterized subprocesses of the cycle. By Lemma 2, any state $x$ represented by the resulting subgraph is a reachable state of a corresponding subgraph of an instance. This forms the base case of the induction. Now consider a consistent subgraph $\mathcal{S}'$ of $\mathcal{S}$ and assume it represents a reachable state set in $\mathcal{G}$. If $\mathcal{S}'$ and $\mathcal{S}$ are the same; we then have the result by assumption. Otherwise, there must exist a state of an output node in $\mathscr{S}'$ and an arc from that state that exists in $\mathscr{S}$ but not in $\mathscr{S}'$. Therefore there exists a consistent subgraph $\mathscr{S}''$ of $\mathscr{S}$ that is formed by adding to $\mathscr{S}'$ a path from that state to that of the input node (and all arcs $(u', v')$ such that $u'$ and $v'$ are states of parameterized subprocesses within that path.) By assumption, any state represented by $\mathscr{S}'$ is reachable within its corresponding instance. Now consider the subgraph $\mathscr{S}''$ that includes the new paths. By the proof of Theorem 1(b), any state represented by this subgraph is reachable. This completes the induction. Therefore state $x$ represented by $\mathscr{S}$ is a reachable state in PCN $\mathcal{G}$.

*(Part (b))* Consider an instance of the PCN that is in a (reachable) total deadlocked state $x$. Since $x$ is reachable, by Proposition 4, the state $x_j$ of any output subprocess $G_j$, is in relation $\mathcal{Q}_{j+1}$ with the state of one of its direct successor subprocesses ($\mathcal{Q}_{j+1}$ is a weak invariant simulation w.r.t. all the shared events of $G_j$ that are not shared with the direct predecessor of $G_j$ in the instance.) Therefore, there exists a cycle in the instance of the PCN such that the states of all output subprocesses of that cycle are in such weak invariant simulation relations with the states of their direct successors in the cycle. Let this cycle be cycle number one. Consider the isolated version of cycle number one. By Proposition 2(a), the state of output subprocess $\hat{G}_j$ of the isolated cycle is in a relation $\mathcal{V}_j$ with the state of its direct successor in the cycle ($\mathcal{V}_j$ is a weak invariant simulation w.r.t. shared events between the two subprocesses). Since $G_j$ was chosen arbitrarily, all output subprocesses of the cycle are in such weak invariant simulation relations. By Proposition 2(b), (5) also holds for this isolated cycle. Assumptions (1, 2, 3, 4) are not affected by the isolation operation. Therefore by the proof of Theorem 1(a), isolated cycle number one is deadlocked if and only if its subprocesses are in states represented by a consistent subgraph of the dependency graph. Let $\hat{\mathscr{S}}$ be such a subgraph.

If no state of any output subprocess of cycle number one has an event enabled from it that is shared with a direct successor for which there is no state in $\hat{\mathscr{S}}$, then $\hat{\mathscr{S}}$ is full. Otherwise, let $G_{j+1}$ be a direct successor of $G_j$ not belonging to cycle number one such that an event shared between $G_{j+1}$ and $G_j$ is enabled from $x_j$. This means that two different events are defined from $x_j$, shared with two different direct successors of $G_j$. By Proposition 7, there exists a string $r \in \Sigma_j^*$ containing only local events and events shared with $G_{j-1}$ such that $x_j \in \xi_j(x_j^0, r)$. Therefore by (6) and the definition of weak invariant simulation, $(x_j, x_{j+1}^0) \in \hat{\mathcal{Q}}'_{j+1}$, where $\hat{\mathcal{Q}}'_{j+1}$ is a weak invariant simulation of $G_{j+1}$ by $G_j$ w.r.t. all the shared events of $G_j$ that are not shared with the direct predecessor of $G_j$. Consider another cycle in the instance that contains the input subprocess and $G_j$ and $G_{j+1}$. Set those components of the total deadlocked state corresponding to subprocesses of this cycle as the new initial states of subprocesses of this cycle and label this new cycle as cycle number two. The isolated version of cycle number two satisfies all assumptions of Theorem 1(a). It is therefore deadlocked if and only if it is in a state represented by a consistent subgraph of the dependency graph. This consistent subgraph is merged with $\hat{\mathscr{S}}$ to form a consistent subgraph representing the states of both cycles. We inductively repeat this procedure for all direct successors of all output subprocesses. The procedure terminates when the consistent subgraph is full. This full, consistent subgraph represents a component of the total deadlocked state. □

**Remark 4.** Part (b) of the above theorem gives only a necessary condition for the existence of a reachable total deadlock in an instance of the PCN. But by part (a), this condition implies the existence of a reachable partial deadlock that includes the input subprocess. Any events that can be executed within an instance whose state includes such a partial deadlock are therefore necessarily restricted to an acyclic subgraph of the instance that does not include the input node. Such behavior would arguably be considered undesirable or pathological in many applications (In our traffic network example, this amounts to trains continually moving back and forth along one of the routes). In such cases, Theorem 1 provides a necessary and sufficient condition for the existence of a reachable total deadlock.

Consider the traffic network example. The dependency graph (Figure 3(a)) of this PCN has four full consistent subgraphs that represent states if its instances. Figure 3 parts (b) and (c) shows two of these subgraphs that represent states of the top loop of the network. Full, consistent subgraphs of this dependency graph represent states of instances of the PCN with odd numbers of spaces in the top or bottom routes (See Remark 3). According to Theorem 2(a) these states are partial deadlocks of the traffic network. In fact, this is one of those cases in which events cannot occur indefinitely in an acyclic subgraph of an instance that does not include the input node. Therefore, when the network is in these states, it will eventually enter a total deadlock. According to Theorem 2(b) the network is free of total deadlock if the lengths of the top and bottom routes are both even.

## VI. Conclusion

The deadlock analysis of a parameterized-chain discrete event network was addressed in this paper. We developed the dependency graph to cover possible interaction scenarios and to verify the potential occurrence of generalized circular waits as formalized via the notion of full, consistent subgraphs of the dependency graph.

We showed that the existence of such a circular wait is a necessary condition for the existence of a reachable total deadlock of an instance of the network, and a sufficient condition for the existence of a reachable partial deadlock that includes the unique 'input' subprocess of the network. Under such a partial deadlock, executable events are confined to an acyclic subgraph that does not contain the input subprocess. In applications in which such behavior cannot occur, the necessary condition for total deadlock becomes a sufficient one. We emphasize that this work relates to parameterized networks – that is, to infinite families of finite-state network instances. Thus, the total state set under consideration is infinite.

Bherer *et al.* have proposed a control synthesis procedure for parameterized networks [15] without addressing blocking issues. Our long-term goal is to develop nonblocking supervisor synthesis methods for tractable subclasses of parameterized networks.

## Appendix

In this section we present all of the intermediate results used in proofs of Theorems 1 and 2.

We first present some properties of PCN defined in our framework. The next proposition expresses that under synchronization of shared events, a weak invariant simulation of $G_i$ by $G_{i-1}$ with respect to shared events between them is preserved.

**Proposition 1.** [2] Consider two arbitrary generators $G_k = (X_k, \Sigma_k, \xi_k, x_k^0)$, $k \in \{i, i+1\}$, and a synchronous product $G_i \| G_{i+1} = (X, \Sigma, \xi, x^0)$. For all $(x_i, x_{i+1}) \in X$, and $\forall s \in \Sigma^*$, we have

$$[(x_i, x_{i+1}) \in \mathcal{V}_{i+1} \ \& \ (x_i', x_{i+1}') \in \xi((x_i, x_{i+1}), s)]$$
$$\Rightarrow (x_i', x_{i+1}') \in \mathcal{V}_{i+1}, \qquad (12)$$

where $\mathcal{V}_{i+1}$ is a weak invariant simulation relation of $G_{i+1}$ by $G_i$ w.r.t. $\Sigma_i \cap \Sigma_{i+1}$.

We imposed restrictions on input and output subprocesses of each cycle by (4-6). The next proposition expresses two properties of input and output subprocesses.

Part (a) of the proposition expresses that whenever states of two neighbors in a cycle are in the weak invariant simulation relation of assumption (6), then these states belong to another weak invariant simulation relation in the isolated cycle; however, part (b) indicates that assumption (5) is essentially preserved for the restricted subprocesses of the isolated cycle.

**Proposition 2.** Consider a cycle $G^N$ and an isolated cycle $\hat{G}^N$.

(a) Let $j \in J$, where $J$ is the index set of output subprocesses of $G^N$, and let $\mathcal{Q}_{j+1}$ be a weak invariant simulation of $G_{j+1}$ by $G_j$ w.r.t. $\Sigma_{S_j} \setminus \Sigma_{j-1}$. Then the restriction of $\mathcal{Q}_{j+1}$ to pairs of states of $\hat{G}_j$ and $\hat{G}_{j+1}$ is a weak invariant simulation of $\hat{G}_{j+1}$ by $\hat{G}_j$ w.r.t. $\Sigma_j \cap \Sigma_{j+1}$.

(b) Let $\hat{R}$ be the state set of the synchronous product $\hat{G}_1 \| \hat{G}_2$. For all $(x_1, x_2) \in \hat{R}$, $(x_1 \hat{\mathcal{W}} x_2)$, where $\hat{\mathcal{W}}$ is a weak invariant simulation of $\hat{G}_2$ by $\hat{G}_1$ w.r.t. all shared events of $G_1$.

*Proof.* (Part (a)) Assume $(x_j, x_{j+1}) \in \mathcal{Q}_{j+1}$. We show that $(x_j, x_{j+1}) \in \hat{\mathcal{V}}_{j+1}$, where $\hat{\mathcal{V}}_{j+1}$ is a weak invariant simulation of $\hat{G}_{j+1}$ by $\hat{G}_j$ w.r.t. $\Sigma_j \cap \Sigma_{j+1}$. Since $(x_j, x_{j+1}) \in \mathcal{Q}_{j+1}$, by the definition of weak invariant simulation, for any $l_{j+1} \in \Sigma_{j+1}^*$ with $\xi_{j+1}(x_{j+1}, l_{j+1}) \neq \emptyset$, there exists string $l_j \in \Sigma_j^*$ such that $\xi_j(x_j, l_j) \neq \emptyset$,

$$P_{\Sigma_{S_j} \setminus \Sigma_{j-1}}(l_j) = P_{\Sigma_{S_j} \setminus \Sigma_{j-1}}(l_{j+1}), \tag{13}$$

and for any $x'_j \in \xi_j(x_j, l_j)$ and $x'_{j+1} \in \xi_{j+1}(x_{j+1}, l_{j+1})$, $(x'_j, x'_{j+1}) \in \mathcal{Q}_{j+1}$. But $l_{j+1} \in \Sigma_{j+1}^*$, therefore

$$P_{\Sigma_{S_j} \setminus \Sigma_{j-1}}(l_{j+1}) = P_{\Sigma_j \cap \Sigma_{j+1}}(l_{j+1}). \tag{14}$$

Therefore by (13), $l_j$ contains no event shared with any other direct successor of $G_j$ in an instance; therefore $\hat{\xi}_j(x_j, l_j) \neq \emptyset$ and $P_{\Sigma_{S_j} \setminus \Sigma_{j-1}}(l_j) = P_{\Sigma_j \cap \Sigma_{j+1}}(l_j)$. By (13) and (14), $P_{\Sigma_j \cap \Sigma_{j+1}}(l_j) = P_{\Sigma_j \cap \Sigma_{j+1}}(l_{j+1})$. Note that because the pair $(x'_j, x'_{j+1})$ is also member of $\mathcal{Q}_{j+1}$, we conclude that the restriction of $\mathcal{Q}_{j+1}$ to pairs of states of $\hat{G}_j$ and $\hat{G}_{j+1}$ is a suitable $\hat{\mathcal{V}}_{j+1}$.

(Part(b)) By the fact that $\hat{R} \subseteq R$, where $R$ is the state set of the synchronous product $G_1 \| G_2$ and the definition of weak invariant simulation. Details are omitted due to similarity to proof of part(a). □

The following proposition expresses that state set of synchronous product of any two neighboring subprocess in an isolated cycle are in weak invariant simulation w.r.t. shared events between them.

**Proposition 3.** Consider cycle $G^N = (X, \Sigma, \xi, x^0)$ of an instance of a PCN satisfying (1-6). For all $(x_i, x_{i+1}) \in \hat{R}_i + 1$, where $\hat{R}_{i+1}$ is the state set of synchronous product $\hat{G}_i \| \hat{G}_{i+1}$,

$$(x_i, x_{i+1}) \in \mathcal{V}'_{i+1}. \tag{15}$$

where $\mathcal{V}'_{i+1}$ is a weak invariant simulation of $\hat{G}_{i+1}$ by $\hat{G}_i$ w.r.t. $\Sigma_i \cap \Sigma_{i+1}$.

*Proof.* By assumption, $(x_i^0, x_{i+1}^0) \in \mathcal{V}_{i+1}$, where $\mathcal{V}_{i+1}$ is a weak invariant simulation of $G_{i+1}$ by $G_i$ w.r.t. $\Sigma_i \cap \Sigma_{i+1}$. But according to the PCN structure only input and output subprocesses are affected by the isolation operation and $\hat{G}_i$ and $G_i$ are the same for $i \notin J \cup \{1\}$, where $J$ is the index set of output subprocesses. The proof for weak invariant simulation of $\hat{G}_{i+1}$ by $\hat{G}_i$ for $i \notin J \cup \{1\}$ follows from Proposition 1 and the fact that $(\Sigma_i \cap \Sigma_{i+1}) \subseteq \Sigma_{i+1}$. For $i \in J$, the proof is by Proposition 2(a). For $i = 1$, we use the result of Proposition 2(b). By this proposition, for all $(x_1, x_2) \in \hat{R}_2$, $(x_1 \hat{\mathcal{W}} x_2)$, where $\hat{\mathcal{W}}$ is a weak invariant simulation of $\hat{G}_2$ by $\hat{G}_1$ w.r.t. all shared events of $G_1$. By definition weak invariant simulation, this implies that $(x_1 \hat{\mathcal{V}}'_2 x_2)$. □

The following lemma expresses an important property of our proposed network: let $\hat{G}^N$ be an isolated cycle of an instance a PCN satisfying (1-6). Then in any reachable state of an instance of the PCN, all the shared events of a given subprocess $\hat{G}_i$, $1 < i \leq N$, with the neighbor of 'lower' index, namely $\hat{G}_{i-1}$, can eventually be executed via a string whose execution does not change the states of subprocesses $\hat{G}_{i+1}$ to $\hat{G}_N$.

**Lemma 1.** Consider cycle $G^N$ of an instance of a PCN satisfying (1-6) and let $\hat{G}^N = (\hat{X}, \Sigma, \hat{\xi}, x^0)$ be the isolated version of $G^N$. For all $x \in \hat{X}$ and all $1 < i \leq N$, we have

$$(\forall \sigma_{i-1} \in \Sigma_{i-1} \cap \Sigma_i)[(\hat{\xi}_i(x_i, \sigma_{i-1}) \neq \emptyset)$$
$$\Rightarrow (\exists s \in (\Sigma \setminus \bigcup_{k=i}^{N} \Sigma_k)^*)(\hat{\xi}(x, s\sigma_{i-1}) \neq \emptyset)], \tag{16}$$

*Proof.* $\hat{G}^N$ has the topology of a ring network. Let $x \in X$ be a global state such that for some $i \neq 1$ and $\sigma_{i-1} \in \Sigma_{i-1} \cap \Sigma_i$,

$$\xi_i(\hat{x}_i, \sigma_{i-1}) \neq \emptyset. \tag{17}$$

According to Proposition 3, $(x_{i-1}, x_i) \in \mathcal{V}'_i$, where $\mathcal{V}'_{i+1}$ is a weak invariant simulation of $\hat{G}_{i+1}$ by $\hat{G}_i$ w.r.t. $\Sigma_i \cap \Sigma_{i+1}$. Therefore by (17) and the definition of weak invariant simulation, there exists a string $l_{i-1} \in (\Sigma_{i-1} \setminus \Sigma_i)^*$ such that $\chi_{i-1}(\sigma_{i-1}) \in \hat{\xi}_{i-1}(x_{i-1}, l_{i-1})$. If $l_{i-1}$ has no events shared with $G_{i-2}$ (consists of local events only); $\chi_{i-1}(\sigma_{i-1})$ can be reached in the global model by a local string of $G_{i-1}$. This satisfies (16).

For the case that $l_{i-1}$ contains shared events with $G_{i-2}$, the proof of reachability of $\chi_{i-1}(\sigma_{i-1})$ in $G_{i-1}$ within the global model is by induction on subprocess indices. To form the base case of the induction, let $i = 2$. By Proposition 2(b), for all $x \in \hat{R}$, $x_1 \hat{\mathcal{W}} x_2$, where $\hat{\mathcal{W}}$ is a weak invariant simulation of $\hat{G}_2$ by $\hat{G}_1$ w.r.t. all shared events of $G_1$ and $\hat{R}$ is the state set of the synchronous product $\hat{G}_1 \| \hat{G}_2$. By (17), we have $\hat{\xi}_2(x_2, \sigma_1) \neq \emptyset$; then, according to the definition of weak invariant simulation, there exists string $l_1 \in (\Sigma_{L_1})^*$ such that $\hat{\xi}_1(x_1, l_1 \sigma_1) \neq \emptyset$. Since $l_1$ consists of only local events of $G_1$, we have $l_1 \in (\Sigma \setminus \bigcup_{j=2}^N \Sigma_j)^*$ and satisfies (16). This forms the base case of the induction. s

For the induction hypothesis, assume (16) holds when $i$ is replaced with some $k > 1$; we will show that (16) holds for $k + 1$. Suppose that for some event $\sigma_k \in \Sigma_k \cap \Sigma_{k+1}$, we have $\hat{\xi}_{k+1}(x_{k+1}, \sigma_k) \neq \emptyset$. Then by the same reasoning as above, there exists a string $l_k \in (\Sigma_k \setminus \Sigma_{k+1})^*$ such that $\chi_k(\sigma_k) \in \hat{\xi}_k(x_k, l_k)$. The only possible shared events of $l_k$ are with $G_{k-1}$. Let $\alpha_{k-1}$ be the first shared event of $l_k$ with $G_{k-1}$. According to the induction hypothesis, there exists string $s \in (\Sigma \setminus \bigcup_{j=k}^N \Sigma_j)^*$ such that $\hat{\xi}(x, s\alpha_{k-1}) \neq \emptyset$. Therefore shared event $\alpha_{k-1}$ can be executed within the global model by a string $s$ that contains no event in $\bigcup_{j=k}^N \Sigma_j$. The proof for reachability of the rest of the shared events of $l_k$ within the global network is similar. Since $(\Sigma \setminus \bigcup_{j=k}^N \Sigma_j) \subseteq (\Sigma \setminus \bigcup_{j=k+1}^N \Sigma_j)$, we conclude that there exists a string $\hat{s} \in (\Sigma \setminus \bigcup_{j=k+1}^N \Sigma_j)^*$ such that $P_{\Sigma_k}(\hat{s}) = l_k$ and $\hat{s}$ can be executed within the global model. Since $\chi_k(\sigma_k) \in \hat{\xi}_k(x_k, l_k)$ and $\hat{\xi}_{k+1}(x_{k+1}, \sigma_k) \neq \emptyset$, $\sigma_k$ can be executed and this completes the proof.

$\square$

The next proposition gives an important property of output subprocesses of a PCN. It expresses that regardless of the evolution of a PCN, the state of any output subprocess weakly invariantly simulates the state of one of its direct successors w.r.t all shared events of the output subprocess that are not shared with its direct predecessor.

**Proposition 4.** Consider an instance of a PCN satisfying (1-6). Let $G_j$ be an arbitrary output subprocess and $G_{j-1}$ be its direct predecessor in the instance. For any reachable state $x$ of this instance of the PCN, there exists a direct successor $G_{j+1}$ of $G_j$, such that

$(x_j, x_{j+1}) \in \mathcal{Q}_{j+1}$, where $x_j$ and $x_{j+1}$ are the states of $G_j$ and $G_{j+1}$, and $\mathcal{Q}_{j+1}$ is a weak invariant simulation of $G_{j+1}$ by $G_j$ w.r.t. all the shared events of $G_j$ that are not shared with $G_{j-1}$. ; and (b) for any direct successor $G_{j+1}$ such that $x_j$ is in-sync with the state $x_{j+1}$ of $G_{j+1}$, we have $(x_j, x_{j+1}) \in \mathcal{Q}_j$, where $\mathcal{Q}_j$ is a weak invariant simulation of $G_{j+1}$ by $G_j$ w.r.t. all the shared events of $G_j$ that are not shared with $G_{j-1}$.

*Proof.* Let $G_k = (X_k, \Sigma_k, \xi_k, x_k^0)$, $k = j-1, j, j+1$. Thus the set of shared events of $G_j$ that are not shared with $G_{j-1}$ is $\Sigma_{S_j} \setminus \Sigma_{j-1}$, where $\Sigma_{S_j}$ is the set of shared events of $G_j$.

Let $s$ label a path from the initial state of the instance of the PCN to state $x$. If string $s$ has no event that belongs to $\Sigma_{S_j} \setminus \Sigma_{j-1}$, the desired property holds, by (6) and the definition of weak invariant simulation. In case the string $s$ contains an event in $\Sigma_{S_j} \setminus \Sigma_{j-1}$, let $\sigma$ be the last such shared event symbol in $s$. Let $G_{j+1}$ be the direct successor of $G_j$ that shares event $\sigma$ with $G_j$. Assume $r\sigma$ is the longest prefix of $s$ ending in $\sigma$, and $r_{j+1}$ is the projection of $r$ onto the alphabet of $G_{j+1}$. Then $\chi_{j+1}(\sigma) \in \xi_{j+1}(x_{j+1}^0, r_{j+1})$, where $\chi_{j+1}(\sigma)$ is the unique companion state of $\sigma$ in $G_{j+1}$ (uniqueness is by (2)). By (6), $(x_j^0, x_{j+1}^0) \in \mathcal{Q}_{j+1}$. By construction, $r_{j+1}$ has no symbols shared with any other direct successors of $G_j$l; so by the definition of weak invariant simulation there must exist a string labeling a path from $x_j^0$ to $\chi_j(\sigma)$ in $G_j$ that contains no event shared with other direct successors of $G_j$ (recall $\chi_j(\sigma)$ is the companion state of $\sigma$ in $G_j$). Therefore $(\chi_j(\sigma), \chi_{j+1}(\sigma)) \in \mathcal{Q}_{j+1}$. This in turn means that for any $\hat{x}_j \in \xi_j(\chi_j(\sigma), \sigma)$ and $\hat{x}_{j+1} \in \xi_{j+1}(\chi_{j+1}(\sigma), \sigma)$, $(\hat{x}_j, \hat{x}_{j+1}) \in \mathcal{Q}_{j+1}$. By assumption, $\sigma$ is the last event symbol in $\Sigma_{S_j} \setminus \Sigma_{j-1}$ that occurs in $s$; therefore, by the definition of weak invariant simulation $(x_j, x_{j+1}) \in \mathcal{Q}_{j+1}$. □

The next proposition establishes properties of a state of a cycle represented by the dependency graph of a PCN. It uses these properties to show that state of any cycle represented by a consistent subgraph of the dependency graph is a forward-dependent state of that cycle.

**Proposition 5.** Consider cycle $G^N = \|_{i=1}^N G_i = (X, \Sigma, \xi, x^0)$ of an instance of a PCN satisfying (1-6), and let $\hat{G}^N = (\hat{X}, \Sigma, \hat{\xi}, x^0)$ be the isolated version of $G^N$. Let $x \in X_1 \times X_2 \times \ldots \times X_N$ be represented by a consistent subgraph of the dependency graph $\mathscr{D}$ and $J$ be the index set of output subprocesses of $G^N$. (a) For any $j \in J$, $(x_j, x_{j+1}) \in \mathcal{Q}_{j+1}$, where $\mathcal{Q}_{j+1}$ is the weak invariant simulation of $G_{j+1}$ by $G_j$ w.r.t. $\Sigma_{S_j} \setminus \Sigma_{j-1}$. (b) For any $x_j$, $j \in J$, there exists an event $\sigma_j \in \Sigma_j \cap \Sigma_{j+1}$ such that $\hat{\xi}_j(x_j, \sigma_j) \neq \emptyset$, where $\hat{\xi}_j$ is the transition function of $\hat{G}_j$. (c) Any such state $x$ is a forward-dependent state of $G^N$.

*Proof.* (a) By assumption $x$ is represented by a consistent subgraph of the dependency graph $\mathscr{D}$. Therefore consider an arc $(x_j, x_{j+1})$, $j \in J$ in the dependency graph belonging to that subgraph. By the construction of the dependency graph and the definition of forward-dependency, $(x_j, x_{j+1})$ is reachable in the synchronous product of $\hat{G}_j \| \hat{G}_{j+1}$, where $\hat{G}_j$ and $\hat{G}_{j+1}$ are the isolated versions of $G_j$ and $G_{j+1}$ in isolated cycle $\hat{G}^N$ (because $G_{j+1}$ cannot be input or output subprocess). Therefore by (6) and the definition of weak invariant simulation,

$$(x_j, x_{j+1}) \in \mathcal{Q}_{j+1} \tag{18}$$

where $\mathcal{Q}_{j+1}$ is the weak invariant simulation of $G_{j+1}$ by $G_j$ w.r.t. $\Sigma_{S_j} \setminus \Sigma_{j-1}$.

(b) By (1) and the fact that $G_j$ and $G_{j+1}$ share events, there exists a string $l_{j+1} \in \Sigma_{j+1}^*$ such that $\xi_{j+1}(x_{j+1}, l_{j+1}) \neq \emptyset$ and $P_{\Sigma_j}(l_{j+1}) \neq \epsilon$. Therefore by (18), there must exist a string $l_j \in \Sigma_j^*$ such that $\xi_j(x_j, l_j) \neq \emptyset$ and $P_{\Sigma_{S_j} \setminus \Sigma_{j-1}}(l_{j+1}) = P_{\Sigma_{S_j} \setminus \Sigma_{j-1}}(l_j)$. By definition of forward-dependence, there is no local event enabled from $x_j$, therefore the first event of $l_j$ is in $\Sigma_j \cap \Sigma_{j+1}$.

(c) In order for $x$ to be a forward dependent state of $G^N$, we have to show the reachability of $(x_1, x_2)$ in $\hat{G}_1 \| \hat{G}_2$ and $(x_{j-1}, x_j)$ in $\hat{G}_{j-1} \| \hat{G}_j$, for $j \leq J$. Note that only input and output subprocesses are affected by isolation of a cycle. Therefore, we only have to show the reachability of $(x_{i-1}, x_i)$ in $\hat{G}_{i-1} \| \hat{G}_i$ for $i \in \{1\} \cup J$. For $i = j \in J$, by part (b), there exists an event $\sigma_j \in \Sigma_j \cap \Sigma_{j+1}$ such that $\hat{\xi}_j(x_j, \sigma_j) \neq \emptyset$. Consider string $k_{j+1} \in \Sigma_{j+1}^*$ such that $\xi_{j+1}(x_{j+1}^0, k_{j+1}\sigma_j) \neq \emptyset$. Then by (6) there exists $k_j \in \Sigma_j^*$ such that $\xi_j(x_j^0, k_j\sigma) \neq \emptyset$ and $P_{\Sigma_{S_j} \setminus \Sigma_{j-1}}(k_{j+1}) = P_{\Sigma_{S_j} \setminus \Sigma_{j-1}}(k_j)$. Therefore $k_j$ contains no event shared with the rest of the direct successors of $G_j$.

On the other hand, by construction of the dependency graph, the pair $(x_{j-2}, x_{j-1})$ is also forward-dependent within some isolated cycle. Therefore, the only events enabled from $x_{j-1}$ are shared with $G_j$. For simplicity assume that $\beta_j$ is the only such event (some such event exists by strong connectivity). Because $(x_{j-1}, x_j)$ is reachable in $G_{j-1} \| G_j$, therefore, by Proposition 1, $(x_{j-1}, x_j) \in \mathcal{V}_j$, where $\mathcal{V}_j$ is a weak invariant simulation w.r.t. $\Sigma_{j-1} \cap \Sigma_j$. Therefore $\beta_j$ is the only event in $\Sigma_{j-1} \cap \Sigma_j$ that is executable from $x_j$ via a string in $(\Sigma_j \setminus \Sigma_{j-1})^*$. By (3), $(x_{j-1}^0, x_j^0) \in \mathcal{V}_j$. Since $x_j$ is reachable from $x_j^0$ by $k_j$ (by assumption

(2)), there must exist a string $k_{j-1}$ labeling a path from $x_{j-1}^0$ to some state $x'_{j-1}$ such that $(x'_{j-1}, x_j) \in \mathcal{V}_j$. Since $\beta_j$ is the only event in $\Sigma_{j-1} \cap \Sigma_j$ executable from $x_j$ via a string in $(\Sigma_j \setminus \Sigma_{j-1})^*$, there must exist a string $s_{j-1}$ labeling a path from $x'_{j-1}$ to a companion state of $\beta_j$ in $G_{j-1}$ such that $s_{j-1}$ contains no event in $\Sigma_{j-1} \cap \Sigma_j$. But by (2), $x_{j-1}$ is the unique companion state of $\beta_j$ in $G_{j-1}$. Therefore $x_{j-1}$ can be reached from $x_{j-1}^0$ by a string $k_{j-1}s_{j-1}$ such that $P_{\Sigma_{j-1} \cap \Sigma_j}(k_{j-1}s_{j-1}) = P_{\Sigma_{j-1} \cap \Sigma_j}(k_j)$. Therefore $(x_{j-1}, x_j)$ is reachable in $\hat{G}_{j-1} \| \hat{G}_j$.

To show the reachability of $(x_1, x_2)$ in $\hat{G}_1 \| \hat{G}_2$, note that because $(x_N, x_1)$ belongs to the dependency graph, the only events enabled from $x_1$ in $\hat{G}_1$ are shared with $G_2$ (note that by (1) some such event exists). By (4), there is no other shared event enabled from $x_1$ in $G_1$. For simplicity assume that $\alpha_1 \in \Sigma_1 \cap \Sigma_2$ is the only event enabled from $x_1$. By (5), $(x_1, x_2) \in \mathcal{W}$, where $\mathcal{W}$ is a weak invariant simulation w.r.t. $\Sigma_{S_1}$. Therefore $\alpha_1$ is the only event in $\Sigma_{S_1}$ that is executable from $x_2$ by a string in $(\Sigma_2 \setminus \Sigma_{S_1})^*$.

By (5), $(x_1^0, x_2^0) \in \mathcal{W}$. Let $k_2$ be a string labeling a path from $x_2^0$ to $x_2$; there must exist a string $k_1$ labeling a path from $x_1^0$ to some state $x'_1$ such that $(x'_1, x_2) \in \mathcal{W}$. Since $\alpha_1$ is executable from $x_2$ by a string in $\Sigma_2 \setminus \Sigma_{S_1}$, there must exist a string $k'_1 \in (\Sigma_1 \setminus \Sigma_{S_1})^*$ labeling a path from $x'_1$ to a companion state of $\alpha_1$ in $G_1$. But by (2), $x_1$ is the unique companion state of $\alpha_1$ in $G_1$. Therefore $(x_1, x_2)$ is reachable in $\hat{G}_1 \| \hat{G}_2$.

Hence, we conclude that $x$ is in fact a forward dependent state of $G^N$. □

We shall show that any state $x_d \in X_d$ that satisfies the forward-dependency property is reachable in $\hat{G}^N$. This in turn means that $x_d$ is reachable in $G^N$. The next proposition is the first step in proving this reachability. It states that in $G^N$, if there exists $x_d \in X_d$ and a reachable state $x \in \hat{X}$ of $\hat{G}^N$ such that for some $k$, $1 \leq k \leq N$, $x_{d_k}$ and $x_k$ are one and the same, then there exists an executable string $l$ that takes $G_{k-1}$ from $x_{k-1}$ to $x_{d_{k-1}}$ and contains no event from alphabets of subprocesses $G_k$ to $G_N$.

**Proposition 6.** Consider cycle $G^N = \|_{i=1}^N G_i = (X, \Sigma, \xi, x^0)$, $1 \leq i \leq N$, of an instance of a PCN satisfying (1-6). Let $\hat{G}^N = \|_{i=1}^N \hat{G}_i = (\hat{X}, \Sigma, \hat{\xi}, x^0)$, $1 \leq i \leq N$, be the isolated version of $G^N$. Consider state $x_d \in X_d$ of Definition 6, and a state $x \in \hat{X}$. For any $k \in \{2, 3, ..., N\}$, if $x_k$ and $x_{d_k}$ are one and the same, then there exists a string $l \in (\Sigma \setminus \bigcup_{r=k}^N \Sigma_r)^*$ such that $\hat{\xi}(x, l) \neq \emptyset$ and $x_{d_{k-1}} \in \hat{\xi}_{k-1}(x_{k-1}, P_{\Sigma_{k-1}}(l))$, where $\hat{\xi}_{k-1}$ is the transition function of $\hat{G}_{k-1}$.

*Proof.* From Remark 2, only the distinguished subprocesses $G_1$ and $G_j$, $j \in J$, where $J$ is the index set of output subprocesses, are affected by the isolation of cycle $G^N$. By (3), for all $i \notin \{1\} \cup J$, we have $(x_i^0, x_{i+1}^0) \in \mathcal{V}_{i+1}$, where $\mathcal{V}_{i+1}$ is a weak invariant simulation of $\hat{G}_{i+1}$ by $\hat{G}_i$ w.r.t. $\Sigma_i \cap \Sigma_{i+1}$. According to Proposition 2(a), $(x_j^0, x_{j+1}^0) \in \mathcal{V}_{j+1}$. By Proposition 2(b) and definition weak invariant simulation, it can easily be shown that $\hat{G}_1 \mathcal{V}_2 \hat{G}_2$. Hence for all $i$, $1 \leq i \leq N$,

$$(x_{i-1}^0, x_i^0) \in \mathcal{V}_i \tag{19}$$

By assumption, $x_d \in X_d$; so we have $(x_{d_{k-1}}, x_{d_k}) \in \hat{R}_k$, where $\hat{R}_k$ is the state set of synchronous product $\hat{G}_{k-1} \| \hat{G}_k$. Therefore, by assumption (3) and Proposition 3, $(x_{d_{k-1}}, x_{d_k}) \in \mathcal{V}_k$; but $x_k$ and $x_{d_k}$ are the same states, hence

$$(x_{d_{k-1}}, x_k) \in \mathcal{V}_k. \tag{20}$$

Then again, $x$ is a reachable state in $\hat{G}^N$, so $(x_{k-1}, x_k) \in \hat{R}_k$, where $\hat{R}_k$ is the state set of synchronous product $\hat{G}_{k-1} \| \hat{G}_k$. Therefore, by (19) and Proposition 3, $(x_{k-1}, x_k) \in \mathcal{V}_k$. By assumption (1) of the network and the definition of $X_d$, there exists a shared event $\beta_{k-1} \in \Sigma_{k-1} \cap \Sigma_k$ that is executable in $\hat{G}_k$ from $x_k = x_{d_k}$ via a string in $(\Sigma_k \setminus \Sigma_{k-1})^*$. The pair $(x_{d_{k-1}}, x_{d_k})$ satisfies the forward-dependency property (7), therefore any transition enabled from $x_{d_{k-1}}$ in $\hat{G}_{k-1}$ is shared with $\hat{G}_k$. Consequently by (20) and the definition of weak invariant simulation, we have

$$\hat{\xi}_{k-1}(x_{d_{k-1}}, \beta_{k-1}) \neq \emptyset. \tag{21}$$

On the other hand, because $\beta_{k-1}$ is accessible from $x_k$ via a string in $(\Sigma_k \setminus \Sigma_{k-1})^*$, and $(x_{k-1}, x_k) \in \mathcal{V}_k$, by the definition of weak invariant simulation there exists a string $\hat{l}_{k-1} \in (\Sigma_{k-1} \setminus \Sigma_k)^*$ and a state $\tilde{x}_{k-1} \in \chi_{k-1}(\beta_{k-1})$ such that

$$\tilde{x}_{k-1} \in \hat{\xi}_{k-1}(x_{k-1}, \hat{l}_{k-1})$$

Therefore by assumption (2) of the network and (21), there exists a $l_{k-1} \in (\Sigma_{k-1} \setminus \Sigma_k)^*$ such that

$$x_{d_{k-1}} \in \hat{\xi}_{k-1}(x_{k-1}, l_{k-1}).$$

However, such an $l_{k-1}$ may contain events shared with $\hat{G}_{k-2}$. According to Lemma 1, for any $i$, $1 < i \leq N$, all these shared events can be executed in the $\hat{G}^N$ by strings with empty projection into $\bigcup_{r=k}^N \Sigma_r$. By repeating this argument, it can be shown that there exists a global string $l \in (\Sigma \setminus (\bigcup_{r=k-1}^{N} \Sigma_r))^*$ such that $P_{\Sigma_{k-1}}(l) = l_{k-1}$ and $l_{k-1}$ can be executed in $G_{k-1}$ within $\hat{G}^N$. □

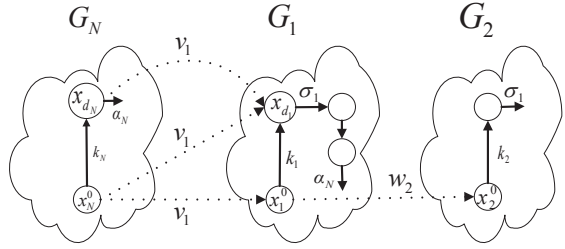

Fig. 5. Subprocesses $G_N$ and $G_1$ and the weak invariant simulation relation between them in the proof of Lemma 2.

Now, using the above proposition, the next lemma shows that any state in the state set of a cycle of our proposed network that satisfies forward-dependency property of Definition 6 is reachable within the isolated cycle, and hence in the global PCN.

**Lemma 2.** Consider cycle $G^N = (X, \Sigma, \xi, x^0)$ of an instance of a PCN satisfying (1-6), and let $\hat{G}^N = (\hat{X}, \Sigma, \hat{\xi}, x^0)$ be the isolated version of $G^N$. All the members of the state set $X_d$ in Definition 6 are reachable in $\hat{G}^N$.

*Proof.* The transitions and weak invariant simulations that appear in this part of the proof are shown in Figure 5. Consider an arbitrary $x_d \in X_d$. For any $i$, $1 \leq i \leq N$, by Definition 6, we have $(x_{d_{i-1}}, x_{d_i}) \in \hat{R}_i$. Therefore by (3) and Proposition 3, we have

$$(\forall i)(x_{d_{i-1}}, x_{d_i}) \in \mathcal{V}_i, \tag{22}$$

where $\mathcal{V}_i$ is a weak invariant simulation w.r.t. $\Sigma_{i-1} \cap \Sigma_i$. By (5) and reachability of $(x_{d_1}, x_{d_2})$ in $G_1 \| G_2$,

$$(x_{d_1}, x_{d_2}) \in \mathcal{W}. \tag{23}$$

where $\mathcal{W}_2$ is a weak invariant simulation w.r.t. $\Sigma_{S_1}$.

According to the definition of isolated cycle $\hat{G}^N$, the only subprocesses affected by isolation are $G_1$ and $G_j$, $j \in J$, where $J$ is the index set of output subprocesses. Therefore $\hat{G}_2$ and

$G_2$ are one and the same. This means that $\hat{G}_2$ satisfies (1). By this assumption, there must exist a path in $\hat{G}_2$ from $x_{d_2}$ to a companion state of an event in $\Sigma_1 \cap \Sigma_2$. Therefore by (23), and the definition of weak invariant simulation, there exists a transition enabled from $x_{d_1}$. But according to (7), the only transitions enabled from $x_{d_1}$ are via events that are shared transitions with $\hat{G}_2$ (if an event enabled from $x_{d_1}$ is a local event or shared event with $G_N$, then $(x_{d_1}, x_{d_2})$ does not satisfy (7)). Let $\sigma_1 \in \Sigma_1 \cap \Sigma_2$ be such that $\xi_1(x_{d_1}, \sigma_1) \neq \emptyset$. By the definition of $X_d$, $\chi_2(\sigma_1)$ is nonempty. Again by (1), there exists a string $k_2 \in \Sigma_2^*$ such that $\chi_2(\sigma_1) \cap \xi_2(x_2^0, k_2) \neq \emptyset$. By (5), $(x_1^0, x_2^0) \in \mathcal{W}_2$; therefore by the definition of weak invariant simulation there exists string $k_1 \in ((\Sigma_1 \cap \Sigma_2) \cup \Sigma_{L_1})^*$ such that $\chi_1(\sigma_1) \cap \xi_1(x_1^0, k_1) \neq \emptyset$. Therefore by (2) and the fact that $k_1 \in ((\Sigma_1 \cap \Sigma_2) \cup \Sigma_{L_1})^*$, we have $x_{d_1} \in \hat{\xi}_1(x_1^0, k_1) \neq \emptyset$. In other words, $x_{d_1}$ is reachable in $\hat{G}_1$. On the other hand, by setting $i = 1$,

$$(x_N^0, x_1^0) \in \mathcal{V}_1. \qquad \text{(by (3))}$$

Therefore by the definition of weak invariant simulation, and the fact that $\hat{k}_1$ contains no event shared with $G_N$,

$$(x_N^0, x_{d_1}) \in \mathcal{V}_1. \tag{24}$$

By (1), there exists a shared event $\alpha_N \in \Sigma_N \cap \Sigma_1$ whose companion state in $G_1$ is accessible from $x_{d_1}$ via strings in $(\Sigma_1 \setminus \Sigma_N)^*$. According to (22), we also have $(x_{d_N}, x_{d_1}) \in \mathcal{V}_1$, and by (7) the only transitions enabled from $x_{d_N}$ are shared transitions with $G_1$. Hence

$$\xi_N(x_{d_N}, \alpha_N) \neq \emptyset. \tag{25}$$

But $(x_N^0, x_{d_1}) \in \mathcal{V}_1$, therefore there must exist a string $k_N \in (\Sigma_N \setminus \Sigma_1)^*$ such that

$$\chi_N(\alpha_N) \in \xi_N(x_N^0, k_N) \qquad \text{(by (2) and (25))}$$

But isolation of $G_N$ has no effect on $G_N$ (because according to network structure, $G_N$ is not an input or output subprocess); therefore $(x_{d_N} \in \hat{\xi}_N(x_N^0, k_N))$. String $k_N$ belongs to the set $(\Sigma_N \setminus \Sigma_1)^*$, in other words it contains no event shared with $G_1$. By Lemma 1, all the shared events of $k_N$ with $\hat{G}_{N-1}$ can eventually be executed, therefore $\hat{G}_N$ can reach $x_{d_N}$ within the global system.

By Proposition 6, $(x_{d_{N-1}}, x_{d_N})$ can be reached in the isolated cycle $\hat{G}^N$. But again, we use Proposition 6 and the fact that $\hat{G}_{N-1}$ can reach $x_{d_{N-1}}$, to show that $x_{d_{N-2}}, x_{d_{N-1}}$, and $x_{d_N}$ are simultaneously reachable in $\hat{G}^N$. With the same reasoning and after $N-1$ times application of Proposition 6, it can be shown that state $x_d \in X_d$ is reachable $\hat{G}^N$. Since $\hat{G}^N$ is the restricted version of $G^N$. □

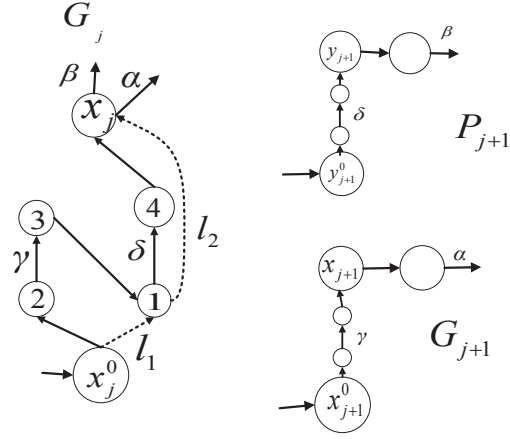

Fig. 6. Subprocesses $G_j$, $G_{j+1}$ and $P_{j+1}$ in the proof of Proposition 7. The unlabeled transitions in $G_j$ contain only local events and events shared with $G_{j-1}$. In order for $x_j, x_{j+1}, y_{j+1}$ to be simultaneously reachable in the global network, there must a path in $G_j$ from state $x_j^0$ to $x_j$ that contains both $\gamma$ and $\delta$. This path is shown by solid transitions. The dotted transition $l_1 l_2$ in $G_j$ consist of local events and events shared with $G_{j-1}$.

The next proposition expresses that in any reachable state of an instance of a PCN satisfying (1-6), for any state $x_j$ of output subprocess $G_j$, if the only events enabled from $x_j$ are shared with direct successors of $G_j$ and at least two events shared with direct successors of $G_j$ are enabled from $x_j$, then state $x_j$ is reachable in $G_j$ by a string containing only local events and events shared with its direct predecessor.

**Proposition 7.** Consider an instance of a PCN satisfying (1-6). Let $G_j$ be an output subprocess, $G_{j-1}$ be its direct predecessor. Consider a reachable state of the PCN instance. In this reachable state, let $x_j$ be a state of $G_j$. Assume the only events enabled from $x_j$ are shared with direct successors of $G_j$ and at least two events shared with direct successors of $G_j$ are enabled from $x_j$. Then there exists a string $r \in \Sigma_j^*$ containing only local events and events shared with $G_{j-1}$ such that $x_j \in \xi_j(x_j^0, r)$.

*Proof.* Assume events $\alpha$ and $\beta$ are enabled from $x_j$ in $G_j$, respectively shared with $G_{j+1}$ and $P_{j+1}$ (two direct successors of $G_j$). Let $G_k = (X_k, \Sigma_k, \xi_k, x_k^0)$, $k = j-1, j, j+1$, and $P_{j+1} = (Y_{j+1}, \Sigma_{P_{j+1}}, \xi_{P_{j+1}}, y_{j+1}^0)$ and $x_{j+1}$ and $y_{j+1}$ be states of $G_{j+1}$, and $P_{j+1}$ in the reachable state.

By (1), and the fact that $\alpha$ is shared with $G_{j+1}$, there exists a string $s_{j+1} \in (\Sigma_{j+1})^*$ such that $\xi_{j+1}(x^0_{j+1}, s_{j+1}\alpha) \neq \emptyset$. If $s_{j+1}$ contains no event shared with $G_j$, then by (6) and (2) such string $r$ exists. Therefore we assume every such $s_{j+1}$ contains an event $\gamma \in \Sigma_j \cap \Sigma_{j+1}$. Similarly, assume any path from $x^0_{P_{j+1}}$ to $\chi_{P_{j+1}}(\beta)$ contains an event $\delta \in \Sigma_j \cap \Sigma_{P_{j+1}}$ (see Figure 6.) By assumption, $x_j$, $x_{j+1}$ and $y_{j+1}$ are simultaneously reachable in the PCN instance. Therefore, there must exist a string $l \in \Sigma_j^*$ such that $x_j \in \xi_j(x_j^0, l)$ and $l$ contains local events and suitable events $\gamma$ and $\delta$ (for the case that $l$ contains multiple events shared with the direct successors of $G_j$, the proof is similar). String $s_{j+1}$ is enabled from $x^0_{j+1}$ in $G_{j+1}$; therefore by (6) there must exist a path from $x_j^0$ to $x_j$ that contains $\delta$ but no event in $\Sigma_j \cap \Sigma_{P_{j+1}}$ (does not contain $\gamma$). With similar reasoning for $P_{j+1}$, there must exist a path from $x_j^0$ to $x_j$ that contains $\delta$ but no event in $\Sigma_j \cap \Sigma_{j+1}$ (hence does not contain $\gamma$). Since by (2) companion states of shared events $\gamma$ and $\delta$ are unique in $G_j$, there exists a string from $x_j^0$ to $x_j$ that contains no event in $\Sigma_j \cap \Sigma_{j+1}$ or $\Sigma_j \cap \Sigma_{P_{j+1}}$. This is demonstrated in Figure 6. Consider the state labellings of this figure for the rest of the proof. By (6), $(x_j^0, x^0_{j+1}) \in \mathcal{Q}_{j+1}$. Therefore by the definition of weak invariant simulation, pair $(3, x_{j+1}) \in \mathcal{Q}_{j+1}$. Since the transition between states 3 and 1 consists only of local events and events shared with $G_{j-1}$, therefore $(1, x_{j+1}) \in \mathcal{Q}_{j+1}$. Since $G_{j+1}$ can reach the companion state of $\alpha$ from $x_{j+1}$, by (2) there must exist path $l_2$ from state 1 to $x_j$, containing only local events and events shared with $G_{j-1}$. On the other hand, by (6) $G_j$ weakly invariantly simulates $P_{j+1}$ w.r.t. $\Sigma_{S_j} \setminus \Sigma_{P_{j-1}}$. Since $\delta$ is reachable from initial state of $P_{j+1}$, by the definition of weak invariant simulation and (2), there must exist a path $l_1$ from initial state of $G_j$ to state 1. Containing only local events and events shared with $G_{j-1}$. Therefore $l_1 l_2$ constitutes a path from initial state of $G_j$ to $x_j$ containing only local events and events shared with $G_{j-1}$. fore $l_1 l_2$ constitutes a path from initial state of $G_j$ to $x_j$ containing only local events and events shared with $G_{j-1}$.

As stated above, for the case that $l$ contains multiple events shared with direct successors of $G_j$, the proof is similar. Here we show the proof sketch for this general case. Let $l$ contain events $\sigma_1 \sigma_2 ... \sigma_n$ shared between $G_j$ and either $G_{j+1}$ or $P_{j+1}$. By (6), $(x_j^0, x^0_{j+1}) \in \mathcal{Q}_{j+1}$ and $(x_j^0, x^0_{j+1}) \in \mathcal{Q}'_{j+1}$, where $\mathcal{Q}_{j+1}$ is a weak invariant simulation of $G_{j+1}$ by $G_j$ and $\mathcal{Q}'_{j+1}$ is a weak invariant simulation of $P_{j+1}$ by $G_j$, both w.r.t. $\Sigma_{S_j} \setminus \Sigma_{j-1}$. Suppose without loss of generality that $\sigma_1 \in \Sigma_j \cap \Sigma_{j+1}$. Then by the definition of weak invariant simulation the unique member of $\chi_j(\sigma_1)$ simulates both $\chi_{j+1}(\sigma_1)$ and $y^0_{j+1}$. It follows that the companion state $\chi_{j+1}(\sigma_2)$ simulates the successor of the first occurrence of $\sigma_1$ and $y^0_{j+1}$; moreover, simulation of $y_{j+1}^0$, there exists a path in $G_j$ from the first companion state to the next that contains no events shared with any direct successor of $G_j$. By replacing this argument for the rest of the path, one can show that any two successive companion states are linked by a path that is not labeled by any events shared with direct successors of $G_j$. It follows that there exists a path from $x_j^0$ to $x_j$ that is labeled only by local events of $G_j$ and events shared with $G_{j-1}$.

□